\documentclass[showpacs,preprint,superscriptaddress,amsmath,amssymb]{revtex4-1}
\usepackage{graphicx}
\usepackage{dcolumn}
\usepackage{bm}
\usepackage{epsfig}
\usepackage{subfigure}
\usepackage{graphics}
\usepackage{amssymb}
\usepackage{amsthm}
\usepackage{amsmath}
\usepackage{array}
\usepackage{color}

\begin{document}

\preprint{APS/123-QED}

r\title{Lattice Boltzmann modeling of wall-bounded ternary fluid flows}

\author{Hong Liang}
 \affiliation{Department of Physics, Hangzhou Dianzi University - Hangzhou 310018, China}

\author{Jiangrong Xu}
 \affiliation{Department of Physics, Hangzhou Dianzi University - Hangzhou 310018, China}

\author{Jiangxing Chen}
 \affiliation{Department of Physics, Hangzhou Dianzi University - Hangzhou 310018, China}

\author{Zhenhua Chai}
\affiliation{School of Mathematics and Statistics, Huazhong University of Science and Technology, Wuhan 430074, China}%
\affiliation{Hubei Key Laboratory of Engineering Modeling and
Scientific Computing, Huazhong University of Science and Technology,
Wuhan 430074,
China}%

\author{Baochang Shi}
\email[Email:~]{shibc@hust.edu.cn.}
\affiliation{School of Mathematics and Statistics, Huazhong University of Science and Technology, Wuhan 430074, China}%
\affiliation{Hubei Key Laboratory of Engineering Modeling and
Scientific Computing, Huazhong University of Science and Technology,
Wuhan 430074,
China}%

\date{\today}

\begin{abstract}
In this paper, a wetting boundary scheme used to describe the
interactions among ternary fluids and solid is proposed in the
framework of the lattice Boltzmann method. This scheme for
three-phase fluids can preserve the reduction consistency property
with the diphasic situation such that it could give physically
relevant results. Combining this wetting boundary scheme and the
lattice Boltzmann (LB) ternary fluid model based on the
multicomponent phase-field theory, we simulated several ternary
fluid flow problems involving solid substrate, including the
spreading of binary drops on the substrate, the spreading of a
compound drop on the substrate, and the shear of a compound liquid
drop on the substrate. The numerical results are found to be good
agreement with the analytical solutions or some available results.
Finally, as an application, we use the LB model coupled with the
present wetting boundary scheme to numerically investigate the
impact of a compound drop on a solid circular cylinder. It is found
that the dynamics of a compound drop can be remarkably influenced by
the wettability of the solid surface and the dimensionless Weber
number.

\end{abstract}
\maketitle

\section{Introduction}
Multiphase flow systems involving ternary fluids and solid substrate
have particular relevance and importance in the fields of
environment and energy, such as enhanced oil recovery~\cite{Maghzi},
proton exchange membrane fuel cell~\cite{hLi}, droplet-based
microfluidic chip~\cite{Seemann}, etc.. Within this context, the
modeling of such flows is a challenging task since it involves the
complex interactions among multiple fluids, and the formation of
multiple contact angles on material substrate. Nonetheless, several
researchers have made a great effort to develop efficient numerical
approaches for simulating ternary fluid flows, which include the
level set method~\cite{Zhao,Saye}, volume of fluid
method~\cite{Bonhomme}, smoothed particle hydrodynamic
method~\cite{Tofighi}, and also the phase field method~\cite{Garcke,
Boyer1, Kim, Boyer2, Said, Dong}. Generally, these traditional
numerical methods based on the macroscopic scale directly solve the
incompressible Navier-Stokes equations coupled with a proper
technique to track the phase interfaces. The methods have their own
impressive versatility in simulating ternary fluid flows, while
similar to two-phase scenario, some of them may have the limitation
more or less, when they are readily applied to interfacial flows
with large topological change~\cite{Anderson}. On the other hand,
the dynamics of fluid interfaces physically can be recognized as a
consequence of intermolecular interactions. In this regard, the
numerical approaches based on the mesoscopic level may be more
suitable to describe complex interfacial dynamics in ternary fluid
systems with or without bounded wall.

The lattice Boltzmann (LB) method~\cite{Guo1}, as a mesoscopic level
method, has received considerable attention in the past two decades.
It has some advantages over the traditional methods such as easy
implementation of complex boundary and high efficiency of code
parallelization. Particularly, due to its kinetic nature, the LB
method can handle fluid-fluid and fluid-solid interactions directly,
which can be regarded as its distinct advantage. From different
physical perspectives, a wide range of multiphase, multicomponent
models have been proposed in the framework of LB method, which can
be commonly divided into four categories: color-gradient
model~\cite{Gunstensen}, pseudo-potential model~\cite{Shan},
free-energy model~\cite{Swift}, and phase-field-based
model~\cite{He, Lee, Liang1, Liang2, Fakhari}. Some improved
variants based on these original multiphase models have also been
proposed, and one can refer to the recent reviews~\cite{Liu, Li1}
and references therein for the detailed expositions. Although a
number of LB models have been developed for the two-phase case shown
above, little attention in comparison has been paid to modelling
multiphase systems involving ternary or more fluids in the LB
community. Lamura {\it{et al.}}~\cite{Lamura} proposed a first
lattice Boltzmann model for oil-water-amphiphile ternary systems,
which is derived based on the minimisation of an appropriate
free-energy functional. However, the model is only suitable to
simulate ternary flows where an amphiphile phase is located at
oil-water interface, and cannot be applied to arbitrary ternary
flows. In addition, none distribution function is introduced to
solely describe the species of amphiphile so that it has no
orientational degree of freedom, which has been revised in the later
developed free-energy model~\cite{QunLi}. Also from the viewpoint of
the free-energy functional, Semprebon {\it{et al.}}~\cite{Semprebon}
recently proposed a LB model for ternary fluids that can adjust
independently the surface tensions among fluids and the contact
angles on the substrate. Chen {\it{et al.}}~\cite{Chen, Nekovee}
developed another lattice Boltzmann model for simulating
oil-water-amphiphile ternary flows, which can be regarded as an
extension of the original pseudo-potential model~\cite{Shan} by
considering interactions among three fluid components. The
generalization of the color-gradient model to multiple immiscible
continuum fluids was attributed to Halliday {\it{et
al.}}~\cite{Halliday1, Halliday2}, who introduced a color gradient
for each of fluid-fluid interfaces in the color model, while their
models are limited to fluids with a very small density difference.
To remove this limitation, Leclaire {\it{et al.}}~\cite{Leclaire}
developed a LB model based on the improved color-gradient model,
where three subcollision operators are also applied. As a result,
their model is able to deal with the multi-componet flows with
moderate density ratios. Recently, Liang {\it{et al.}}~\cite{Liang3}
presented an alternative LB ternary model based on the Cahn-Hilliard
phase-field theory, which provides a firm physical foundation on the
dynamics of the interfaces among three fluids. Actually, the
phase-field based LB models for multiphase flows have showed great
potential in the study of complex interfacial flows~\cite{Liang1,
Liang4}.

As reviewed above, most of the aforementioned LB models only focus
on ternary fluid flows in the absence of bounded solid wall, with a
recent exception~\cite{Semprebon}. Oftentimes, ternary fluids are
encountered with solid substrate in applications mentioned above,
and its wettability plays a vital role in fluid interfacial
dynamics. Therefore, how to describe the interactions among fluids
and solid is a very cruial problem. Our main focus in this paper
will be on the phase-field-based LB ternary model~\cite{Liang3}. As
a continuous work, a suitable wetting boundary scheme that describe
the interations among fluids and solid is proposed in the framework
of the LB method. One distinct feature of the scheme lies in the
reduction consistency property which matches that of the LB ternary
model~\cite{Liang3}. Besides, multiple equilibrium contact angles
can be given explicitly in the boundary condition formulation. The
rest of the paper is organized as follows. In Sec.~\ref{sec:method},
we firstly gives a brief introduction of the LB ternary method, and
then present a novel wetting boundary scheme for ternary fluids.
Numerical experiments to validate the present scheme can be found in
Sec.~\ref{sec:Results}, where a compound drop impact on the solid
cylinder is also studied. At last, we made a summary in
Sec.~\ref{sec: sum}.


\section{LB method for wall-boundary ternary fluid flows}\label{sec:method}

\subsection{LB method for ternary fluid flows}
In this subsection, we give a brief introduction on the LB method
for ternary fluid flows, and a detailed description can be found in
Ref.~\cite{Liang3}. The LB method consists of three LB equations,
two of which is used to capture the interfaces among three-component
fluids and the other is used to derive the fluid velocity and
pressure. The LB evolution equations with the BGK collision operator
can be written as~\cite{Liang3, Guo2}
\begin{subequations}
\begin{equation}
{f^i_k}({\bf{x}} + {{\bf{e}}_k}{\delta_t},t + {\delta_t}) -
{f^i_k}({\bf{x}},t) =  - \frac{1}{\tau_i}\left[ {{f^i_k}({\bf{x}},t)
- f_k^{i,eq}({\bf{x}},t)} \right] + {\delta_t}{F^i_k}({\bf{x}},t),
\end{equation}
\begin{equation}
{g}_k({\bf{x}} +{{\bf{e}}_k}{\delta _t},t + {\delta _t}) -
{g}_k({\bf{x}},t) = -\frac{1}{\tau_g}[{{g}_k({\bf{x}},t) -
g_k^{eq}({\bf{x}},t)}]+ {\delta _t}G_k({\bf{x}},t),
\end{equation}
\end{subequations}
where the superscript $i$ taking $1$ or $2$ represents the $i$-th
phase, $f^i_k({\bf{x}},t)$ and $g_k({\bf{x}},t)$ are the
distribution functions, $f_k^{i,eq}({\bf{x}},t)$ and
$g_k^{eq}({\bf{x}},t)$ are the corresponding equilibrium functions,
$\tau_i$ and ${\tau_g}$ are the non-dimensional relaxation times,
$\delta_t$ is the time step. To recover the macroscopic equations
exactly, the equilibrium distribution functions $f_{k}^{i,eq}$ and
$g_k^{eq}$ are delicately designed as~\cite{Shi, Chai, Liang1}
\begin{subequations}
\begin{equation}
f_{k}^{i,eq} =\left\{
\begin{array}{ll}
 c_i  + ({\omega_k} - 1)\eta {\mu_i},                                       & \textrm{ $k=0$}   \\
 {\omega_k}\eta{\mu_i}  + {\omega_k}c_i{{{\textbf{e}_k} \cdot {\bf{u}}} \over {c_s^2}}, & \textrm{ $k\neq0$},
\end{array}
\right.
\end{equation}
\begin{equation}
g_k^{eq}=\left\{
\begin{array}{ll}
{p \over {c_s^2}}({\omega _k} - 1) + \rho{s_k}({\bf{u}}),              & \textrm{ $k=0$},    \\
{p \over {c_s^2}}{\omega _k} + \rho{s_k}({\bf{u}}),                    & \textrm{ $k\neq0$}, \\
\end{array}
\right.
\end{equation}
\end{subequations}
where $c_i$ is the order parameter that represents the volume
fraction of $i$-th phase within the mixture. In the phase-field
models, one should use three order parameters marked by $c_1$,
$c_2$, and $c_3$ to describe a ternary system, and they are linked
through the constraint~\cite{Boyer1, Kim, Boyer2},
\begin{equation}
 c_1+c_2+c_3=1.
\end{equation}
 In Eqs. (2a) and (2b), $\omega_k$ is the weighting coefficient, $\textbf{e}_k$ is the discrete velocity,
 $c_s$ is the sound speed, $\eta$~is an adjustable parameter, and ${s_k}({\bf{u}})$ is defined
 by~\cite{Liang1,Liang3}
\begin{equation}
{s_k}({\bf{u}}) = {\omega _k}\left[
{{{{{\bf{e}}_k} \cdot {\bf{u}}} \over {c_s^2}} + {{{{({{\bf{e}}_k}
\cdot {\bf{u}})}^2}} \over {2c_s^4}} - {{{\bf{u}} \cdot {\bf{u}}}
\over {2c_s^2}}} \right].
\end{equation}
${\mu_i}$ in Eq. (2a) is the chemical potential, which depends on
the variational derivative of the bulk free energy with respect to
the order parameters in the ternary phase-field models. Up to now,
several researchers have conducted theoretical analyses on the form
of the bulk free energy~\cite{Garcke, Boyer1, Kim, Boyer2, Said}.
Here the one reported in Ref.~\cite{Boyer1, Boyer2} is used since it
can be well-posed and also satisfies the algebraically and
dynamically consistency conditions. Then, the bulk free energy takes
the following form~\cite{Boyer1, Boyer2},
\begin{equation}
F(c_1,c_2,c_3)=\frac{\lambda_1}{2}{c_1^2}(1-c_1)^2+\frac{\lambda_2}{2}{c_2^2}(1-c_2)^2+\frac{\lambda_3}{2}{c_3^2}(1-c_3)^2+{\lambda}{c_1^2}{c_2^2}{c_3^2},
\end{equation}
where ${\lambda}$ is a non-negative parameter, and the chemical
potential ${\mu_i}$ can be derived by~\cite{Boyer1, Boyer2}
\begin{equation}
\mu_i=\frac{4\lambda_T}{D}{\sum\limits_{j\neq{i}}}\left[\frac{1}{\lambda_j}\left(\frac{\partial
F}{\partial{c_i}}-\frac{\partial
F}{\partial{c_j}}\right)\right]-\frac{3}{4}{D}{\lambda_i}\nabla^2{c_i},
\end{equation}
where the parameters ${\lambda_i}$ $(i=1,2,3)$ are related to the surface tensions,
\begin{eqnarray}
{\lambda_1}=\sigma_{12}+\sigma_{13}-\sigma_{23}\nonumber\\
{\lambda_2}=\sigma_{12}+\sigma_{23}-\sigma_{13}\nonumber\\
{\lambda_3}=\sigma_{13}+\sigma_{23}-\sigma_{12},
\end{eqnarray}
where $\sigma_{23}$, $\sigma_{23}$ and $\sigma_{23}$ represent the surface tension between
two fluids of a three-phase system.
 When ${\lambda_i}$ $(i=1,2,3)$ are all positive and further satisfy ${\lambda_i}>0.5{\lambda_T}$ [see Eq. (9)], the bulk free energy with ${\lambda}=0$
can give physically relevant results~\cite{Boyer1, Liang3}, which
will be adopted in our numerical simulations. In this case, one can
simplify Eq. (6) as
\begin{equation}
\mu_i=\frac{12}{D}\left[{\lambda_i}{c_i}(1-c_i)(1-2c_i)-2{\lambda_T}{c_1}{c_2}(1-c_1-c_2)
\right]-\frac{3}{4}{D}{\lambda_i}\nabla^2{c_i},
\end{equation}
where $D$ is the interface thickness, and ${\lambda_T}$ is defined by
\begin{equation}
\frac{3}{\lambda_T}={\sum\limits_{i=1}^3{\frac{1}{\lambda_i}}}.
\end{equation}

In the present work, the D2Q9 lattice model is used without loss of
generality, where the weighting coefficients $\omega_k$ are given by
$\omega_0=4/9$, $\omega_{1-4}=1/9$, $\omega_{5-8}=1/36$, and the
discrete velocities $\textbf{e}_k$ are~\cite{Guo2}
\begin{equation}
\mathbf{e}_{k}=\left\{
\begin{array}{ll}
 (0,0)c,                                                         & \textrm{ $k=0$},   \\
 (\cos [(k-1)\pi /2],\sin [(k-1)\pi /2])c,                       & \textrm{ $k=1-4$}, \\
 \sqrt{2}(\cos [(k-5)\pi /2+\pi /4],\sin [(k-5)\pi /2+\pi /4])c, & \textrm{ $k=5-8$},
\end{array}
\right.
\end{equation}
where $c=\delta_x$/$\delta_t$ is the lattice speed with $\delta_x$
representing the grid spacing, $c_s=c/\sqrt{3}$. For
simplicity, we set the grid space and time increment as the
length and time units, i.e., $\delta_x=\delta_t=1$.

To derive the correct governing equations, the proper source term
${F^i_k}$ and forcing term ${{G}_k}$ should be incorporated in the
LB evolution equation, which can be defined as~\cite{Liang1, Liang3}
\begin{subequations}
\begin{equation}
{{F}^i_k} =(1-\frac{1}{2\tau_i}){{{\omega_k}{{\bf{e}}_k} \cdot {\partial _t}c_i{\bf{u}}} \over {c_s^2}},~~~i=1,~2,
\end{equation}
\begin{equation}
{{G}_k}=(1-\frac{1}{2\tau_g}){({{\bf{e}}_k} - {\bf{u}})} \cdot \left[
{{s_k}({\bf{u}})\nabla{\rho}+({s_k}({\bf{u}})+\omega_k){\frac{({{\bf{F}}_s} + {\bf{G}})}{c_s^2}}}\right]+\frac{{\omega_k}{\mathbf{e}_k}\cdot{\mathbf{F}_a}}{c_s^2},
\end{equation}
\end{subequations}
where ${\bf{G}}$ is the body force, ${{\bf{F}}_a}$ is the additional
interfacial force, ${\bf{F}}_s$ is the surface tension force, which
can take several different forms. Here we take the potential form
$\mathbf{F}_s=\sum\limits_{i=1}^3{\mu_i}\nabla{c_i}$, as widely used
in the ternary phase-field models~\cite{Boyer1, Boyer2, Liang3}.
${{\bf{F}}_a}$ introduced in Eq. (11b) is used to recover the
correct momentum equation, which can be defined as~\cite{Liang3}
\begin{equation}
{{\bf{F}}_a}=\mathbf{u}\sum_{i=1}^2(\rho_i-\rho_3){M_i}\nabla^2\mu_i,
\end{equation}
where $M_i=M_0/\lambda_i$ is the diffusion coefficient in the
interfacial governing equation, and $M_0$ is a positive parameter.
In the LB algorithm, the macroscopic quantities, $c_i$, ${\bf{u}}$
and $p$ are evaluated as~\cite{Liang3},
\begin{subequations}
\begin{equation}
c_i = \sum\limits_k {{f^i_k}},~~~i=1,~2,
\end{equation}
\begin{equation}
{\bf{u}} = \frac{1}{\rho}\left[{{\sum\limits_k {{{\bf{e}}_k}{g_k}} +
0.5{\delta _t}({{\bf{F}}_s} + {\bf{G}})}}\right],
\end{equation}
\begin{equation}
 p = {{c_s^2} \over {(1 - {\omega
_0})}}\left[ {\sum\limits_{k \ne 0} {{g_k}}  + {{{\delta _t}}
\over 2}{\bf{u}} \cdot \nabla \rho + \rho {s_0}(\textbf{u})}
\right],
\end{equation}
\end{subequations}
and the order parameter $c_3$ can be derived from the conservation
(3). For the sake of simplicity, we assume that the fluid density
and viscosity are the linear interpolations of three order
parameters~\cite{Kim}
\begin{equation}
\rho  ={c_1}{\rho_1} + {c_2}{\rho_2} + (1-{c_1}-{c_2}){\rho_3},
\end{equation}
\begin{equation}
\nu  ={c_1}{\nu_1} + {c_2}{\nu_2} + (1-{c_1}-{c_2}){\nu_3},
\end{equation}
where $\rho_i$ and $\nu_i$ $(i=1,2,3)$ are the density and viscosity
of the $i$-th phase. Through Chapman-Enskog analysis~\cite{Liang1,
Liang3}, it is shown that the multi-component Cahn-Hilliard
equations
\begin{equation}
{{\partial{c_i}}\over{\partial t}} + \nabla \cdot{c_i}{\bf{u}}=\nabla\cdot\left({M_i}\nabla {\mu_i}\right),
\end{equation}
and the incompressible Navier-Stokes equations
\begin{subequations}
\begin{equation}
\nabla  \cdot {\bf{u}} = 0,
\end{equation}
\begin{equation}
 \rho ({{\partial {\bf{u}}} \over
{\partial t}} + {\bf{u}} \cdot \nabla {\bf{u}}) =  - \nabla p +
\nabla  \cdot \left[ {\nu\rho(\nabla {\bf{u}} + \nabla
{{\bf{u}}^T})} \right] + {{\bf{F}}_s} + {\bf{G}},
\end{equation}
\end{subequations}
can be derived from the present model. Additionally, one can derive
the expressions of the mobility $M_i$ and the kinematic viscosity
$\nu$ as~\cite{Liang1, Liang3},
\begin{subequations}
\begin{equation}
{M_i} =\eta c_s^2(\tau_i-0.5)\delta t,~i=1,~2,
\end{equation}
\begin{equation}
\nu = c_s^2(\tau_g - 0.5){\delta _t}.
\end{equation}
\end{subequations}
For the numerical computations, the time derivative in Eq. (11a) and
the spatial gradients in Eq. (11b) should be discretized with
suitable difference schemes. In this work, the explicit Euler
scheme~\cite{Shi},
\begin{equation}
\partial_t\chi({\bf{x}},t)=\frac{\chi({\bf{x}},t)-\chi({\bf{x}},t-\delta_t)}{\delta_t},
\end{equation}
is applied for calculating the time derivative and the second-order
isotropic central schemes~\cite{Liang1},
\begin{subequations}
\begin{equation}
 \nabla \chi({\bf{x}},t)=\sum\limits_{k \ne 0}
{\frac{\omega_k\textbf{e}_k\chi({\bf{x}} +{{\bf{e}}_k}{\delta
_t},t)}{c_s^2 \delta_t}}
\end{equation}
\begin{equation}
\nabla^2\chi({\bf{x}},t)=\sum\limits_{k \ne 0}
{\frac{2\omega_k[\chi({\bf{x}} +{{\bf{e}}_k}{\delta
_t},t)-\chi({\bf{x}},t)]}{c_s^2 \delta_t^2}},
\end{equation}
\end{subequations}
are adopted to compute the gradient operators, where $\chi$ represents an arbitrary variable.

\subsection{Wetting boundary condition for ternary fluid flows}

The lattice Boltzmann model for ternary fluid flows is developed
based on the ternary phase-field theory~\cite{Boyer1, Boyer2}, where
the wall wetting effect has not been considered. In order to
simulate three-phase flows in contact with solid wall, a suitable
wetting boundary condition should be established to describe the
interactions among fluids and solid, and its scheme in the framework
of the LB method should also be given. The wetting boundary
condition for ternary fluid flows can be constructed by considering
an additional wall free energy. Denoting the flow domain by $\Omega$
and the solid boundary by ${\partial \Omega}$, the total free energy
of a three-phase system can be expressed as~\cite{Yshi}
\begin{equation}
\Psi_{tot}=\int_\Omega \left[\frac{12}{D}F(c_1, c_2,
c_3)+\sum\limits_{i=1}^3
{\frac{3}{8}D{\lambda_i}|\nabla{c_i}|^2}\right] d\Omega +\int_{\partial \Omega} {\psi_s}(c_1,c_2,c_3)ds,
\end{equation}
where $F(c_1, c_2, c_3)$ is the bulk free energy,
${\psi_s}(c_1,c_2,c_3)$ is the free energy density on the solid
boundary. Boyer {\it{et al.}}~\cite{Boyer1, Boyer2} have showed that
the model without including the boundary effect is algebraically
consistent with the diphasic system only if the bulk free energy
$F(c_1, c_2, c_3)$ and the physical parameters ${\lambda_i}$ are
given by Eqs. (5) and (7), respectively. The expression of the free
energy density ${\psi_s}(c_1,c_2,c_3)$ is also determined based on
the reduction consistency condition. For the convenience of
discussion, we first give a brief overview of the phase-field model
with the wall effect for two-phase flows. In a two-phase system, the
total free energy has the following form~\cite{Lee, Zhang},
\begin{equation}
\Psi=\int_\Omega \left[\frac{12}{D}{\sigma} c^2(1-c)^2+\frac{3}{4}{\sigma} D|\nabla c|^2\right] d\Omega +\int_{\partial \Omega} {\psi_s}(c)ds,
\end{equation}
where $c$ is the order parameter with the values of $0$ and $1$ in
the bulk phase regions, and varies continuously across the
interfacial zone with the thickness $D$, $\sigma$ is the surface
tension between two fluids, ${\psi_s}(c)$ is the free energy density
on the solid wall given by~\cite{Lee, Zhang}
\begin{equation}
{\psi_s}(c)= \sigma_{w1}+(\sigma_{w2}-\sigma_{w1})(3c^2-2c^3),
\end{equation}
where $\sigma_{w1}$ and $\sigma_{w2}$ denote the fluid-wall surface tensions. Minimizing the total free energy, one
can derive the two-phase wetting boundary condition,
\begin{equation}
\frac{3}{2}\sigma D \mathbf{n}\cdot\nabla{c}+\frac{\partial {\psi_s}(c)}{\partial c}=0,
\end{equation}
where $\mathbf{n}$ is the unit normal vector with the direction from
the fluid toward the solid. Substituting Eq. (23) into the above
relation, one can rewrite Eq. (24) as
\begin{equation}
\frac{3}{2}\sigma D \mathbf{n}\cdot\nabla{c}+6(\sigma_{w2}-\sigma_{w1})(c-c^2)=0,
\end{equation}
For the two-phase fluids on the chemically homogeneous wall, the
wettability of the wall can be evaluated by the contact angle
($\theta$), which is determined by the Young's equation associated
with the surface tensions at the fluid-solid junction~\cite{Said,
Dong}, $\cos{\theta}=(\sigma_{w1}-\sigma_{w2})/\sigma$. Then, Eq.
(25) can be recast as
\begin{equation}
\mathbf{n}\cdot\nabla{c}=\frac{4\cos\theta}{D}(c-c^2).
\end{equation}
To be consistent with the diphasic case, the free energy density
${\psi_s}(c_1,c_2,c_3)$ in a three-phase system can be chosen as,
\begin{equation}
{\psi_s}(c_1,c_2,c_3) = \sigma_{w1} (3c_1^2-2c_1^3)+ \sigma_{w2}(3c_2^2-2c_2^3)+\sigma_{w3}(3c_3^2-2c_3^3),
\end{equation}
where $\sigma_{wi}$ $(i=1, 2, 3)$ represents the surface tension
between the solid wall and the $i$-th fluid. One can easily find
that Eq. (27) can reduce to the two phase formulation (23), when
setting $c_1=1-c$, $c_2=c$, $c_3=0$. In the following, we use the
symbol ${\psi_s}$ to mark ${\psi_s}(c_1,c_2,c_3)$ for simplicity.
The wetting boundary condition for ternary fluid flows can be
derived by minimizing the total free energy (21). However, in order
to satisfy the conservation (3), an additional term $\Lambda_i$ as a
function of the order parameters is also introduced. Then, the
wetting boundary condition can be expressed as,
\begin{equation}
\frac{3}{4}{\lambda_i} D \mathbf{n}\cdot\nabla{c_i}+\frac{\partial {\psi_s}}{\partial c_i}+\Lambda_i=0,~i=1,2,3,
\end{equation}
which can be further written as
\begin{equation}
\frac{3}{4}{\lambda_i} D \mathbf{n}\cdot\nabla{c_i}+g_i=0,~i=1,2,3,
\end{equation}
where $g_i$ is defined by $g_i=\frac{\partial {\psi_s}}{\partial
c_i}+\Lambda_i$. Now we give details on how to derive the expression
of $g_i$. Summing Eq. (29) over $i$ and denoting $S=c_1+c_2+c_3$,
one can easily obtain the following equation,
\begin{equation}
\frac{3}{4} D \mathbf{n}\cdot\nabla{S}+\sum_{i=1}^3\frac{g_i}{{\lambda_i}}=0.
\end{equation}
Because of the conservation (3), $S=1$ should be the solution of Eq.
(30), and we can then derive
\begin{equation}
\sum_{i=1}^3\frac{g_i}{{\lambda_i}}=0.
\end{equation}
As pointed in Refs.~\cite{Boyer1, Boyer2}, to be algebraically
consistent with the diphasic system, the ternary model should
preserve the property that the $i$-th phase does not appear during
the time evolution of the system if it is absent at initial time. To
satisfy this property, from Eq. (31) one can obtain the following
relations,
\begin{equation}
{g_i}|_{(c_i=0)}=0,
\end{equation}
for $i$=1, 2 and 3. Supposing ${\Lambda_i}$ being the linear
combination of $\frac{\partial {\psi_s}}{\partial c_j}$~\cite{Yshi},
${g_i}$ can then be written in the vector form,
\begin{equation}
{\mathbf{g}}=\mathbf{A}\cdot\frac{\partial \psi_s}{\partial{\mathbf{c}}},
\end{equation}
where $\mathbf{g}=(g_1,g_2,g_3)^T$, $\mathbf{A}$ is a $3\times3$
matrix, and $\frac{\partial
\psi_s}{\partial{\mathbf{c}}}=(\frac{\partial
\psi_s}{\partial{c_1}}, \frac{\partial \psi_s}{\partial{c_2}},
\frac{\partial \psi_s}{\partial{c_3}})^T$. From the above constraint
conditions, we can choose the matrix $\mathbf{A}$ as
\begin{equation}
\mathbf{A}= \frac{1}{2}\left[
\begin{array}{ccccccc}
\frac{\lambda_1}{\sigma_{12}} \frac{c_2}{1-c_1}+\frac{\lambda_1}{\sigma_{13}} \frac{c_3}{1-c_1}& -\frac{\lambda_1}{\sigma_{12}} \frac{c_1}{1-c_2} & -\frac{\lambda_1}{\sigma_{13}} \frac{c_1}{1-c_3}   \\
-\frac{\lambda_2}{\sigma_{12}} \frac{c_2}{1-c_1} & \frac{\lambda_2}{\sigma_{12}} \frac{c_1}{1-c_2}+\frac{\lambda_2}{\sigma_{23}} \frac{c_3}{1-c_2} & -\frac{\lambda_2}{\sigma_{23}} \frac{c_2}{1-c_3}   \\
-\frac{\lambda_3}{\sigma_{13}} \frac{c_3}{1-c_1} & -\frac{\lambda_3}{\sigma_{23}} \frac{c_3}{1-c_2} & \frac{\lambda_3}{\sigma_{13}} \frac{c_1}{1-c_3}+\frac{\lambda_3}{\sigma_{23}} \frac{c_2}{1-c_3}
\end{array} \right ],
\end{equation}
and $g_i$ can then be derived as
\begin{eqnarray}
 g_1=\frac{3\lambda_1}{\sigma_{13}}(\sigma_{w1}-\sigma_{w3}){c_1}{c_3}+\frac{3\lambda_1}{\sigma_{12}}(\sigma_{w1}-\sigma_{w2}){c_1}{c_2}, \nonumber\\
g_2=\frac{3\lambda_2}{\sigma_{12}}(\sigma_{w2}-\sigma_{w1}){c_1}{c_2}+\frac{3\lambda_2}{\sigma_{23}}(\sigma_{w2}-\sigma_{w3}){c_2}{c_3},\nonumber\\
 g_3=\frac{3\lambda_3}{\sigma_{23}}(\sigma_{w3}-\sigma_{w2}){c_2}{c_3}+\frac{3\lambda_3}{\sigma_{13}}(\sigma_{w3}-\sigma_{w1}){c_1}{c_3}.
\end{eqnarray}
With some algebraic manipulations, one can easily find that Eqs.
(31) and (32) can be satisfied. Additionally, the wetting boundary
condition for three-phase flows given in Eqs. (29) and (35) can
exactly degenerate to the two-phase case when one component
vanishes. For instance, when the $3$-th phase is not present in the
system, i.e., $c_3=0$, and $c_1+c_2=1$, $g_i$ can be simplified as
\begin{eqnarray}
 g_1=\frac{3\lambda_1}{\sigma_{12}}(\sigma_{w1}-\sigma_{w2}){c_1}{c_2},~~g_2=\frac{3\lambda_2}{\sigma_{12}}(\sigma_{w2}-\sigma_{w1}){c_1}{c_2},~~
 g_3=0,
\end{eqnarray}
and the boundary scheme is
\begin{equation}
\frac{3}{2} \sigma_{12} D \mathbf{n}\cdot\nabla{c_2}+3(\sigma_{w2}-\sigma_{w1}){c_1}{c_2}=0,
\end{equation}
which is consistent with the two-phase formulation (25). Considering
the ternary fluids in contact with the chemically homogeneous
substrate, we could describe the wettability of the substrate in
terms of three static contact angles, which satisfy the Young's
relation~\cite{Ding2, Dong},
\begin{equation}
\cos{\theta_{ij}}=\frac{\sigma_{wj}-\sigma_{wi}}{\sigma_{ij}},~~1\leq i<j\leq3,
\end{equation}
where $\theta_{ij}$ is the static contact angle between the wall and
the interface formed by fluids $i$ and $j$. The values of
$\theta_{12}$, $\theta_{13}$, and $\theta_{23}$ cannot be
arbitrarily chosen, and should satisfy the following constraint,
\begin{equation}
\sigma_{12}\cos{\theta_{12}}-\sigma_{13}\cos{\theta_{13}}+\sigma_{23}\cos{\theta_{23}}=0.
\end{equation}
With the substitution of Eq. (35) into Eq. (29) and using the
relation (38), we can ultimately derive the wetting boundary
condition for three-phase flows,
\begin{eqnarray}
 \mathbf{n}\cdot\nabla{c_1}=\frac{4}{D}(\cos{\theta_{13}}{c_1}{c_3}+\cos{\theta_{12}}{c_1}{c_2}), \nonumber\\
 \mathbf{n}\cdot\nabla{c_2}=\frac{4}{D}(-\cos{\theta_{12}}{c_1}{c_2}+\cos{\theta_{23}}{c_2}{c_3}), \nonumber\\
 \mathbf{n}\cdot\nabla{c_3}=\frac{4}{D}(-\cos{\theta_{23}}{c_2}{c_3}-\cos{\theta_{13}}{c_1}{c_3}).
\end{eqnarray}

We now introduce how the three-phase wetting boundary condition is
implemented in the framework of the LB method. The wetting boundary
condition given in Eq. (40) is valid at equilibrium, and thus is
only imposed for the term related to free energy, i.e., $\nabla ^2
c_i$ in Eq. (8). Once the $\nabla ^2 c_i$ is prescribed, $\mu_i$ in
Eq. (8) is treat as a scalar~\cite{Lee}. The $\nabla ^2 c_i$ can be
computed by Eq. (20b). While for the fluid node next to the solid
wall, the computation of $\nabla ^2 c_i$ should be specifically
treated by imposing it the wetting boundary formulation (40), and
the details are given as follows. As depicted in Fig. 1,
$\mathbf{x}_f$ is the fluid node next to the boundary layer,
$\mathbf{x}_w$ is the solid boundary node with one half lattice
length from $\mathbf{x}_f$, and $\mathbf{x}_f+\mathbf{e}_k \delta_t$
is the ghost node. To determine the value of $\nabla ^2 c_i$ at the
$\mathbf{x}_f$, the macroscopic information at the ghost node
$\mathbf{x}_f+\mathbf{e}_k \delta_t$ should be specified. After the
central discretization for the left-hand side of Eq. (40), we then
get
\begin{eqnarray}
\frac{{c_{{1},\mathbf{x}_f+\mathbf{e}_k\delta_t}}-{c_{{1},\mathbf{x}_f}}}{\delta_x}=\frac{4}{D}(\cos{\theta_{13}}{c_{1,\mathbf{x}_w}}{c_{3,\mathbf{x}_w}}
+\cos{\theta_{12}}{c_{1,\mathbf{x}_w}}{c_{2,\mathbf{x}_w}}), \nonumber\\
\frac{{c_{{2},\mathbf{x}_f+\mathbf{e}_k\delta_t}}-{c_{{2},\mathbf{x}_f}}}{\delta_x}=\frac{4}{D}(-\cos{\theta_{12}}{c_{1,\mathbf{x}_w}}{c_{2,\mathbf{x}_w}}
+\cos{\theta_{23}}{c_{2,\mathbf{x}_w}}{c_{3,\mathbf{x}_w}}), \nonumber\\
\frac{{c_{{3},\mathbf{x}_f+\mathbf{e}_k\delta_t}}-{c_{{3},\mathbf{x}_f}}}{\delta_x}=\frac{4}{D}(-\cos{\theta_{23}}{c_{2,\mathbf{x}_w}}{c_{3,\mathbf{x}_w}}
-\cos{\theta_{13}}{c_{1,\mathbf{x}_w}}{c_{3,\mathbf{x}_w}}).
\end{eqnarray}
As shown above, the variables ${c_{{i},\mathbf{x}_w}}$ $(i=1,2,3)$
that represent the distributions of the phase fields at the solid
wall are unknown. Here we use the interpolation
${c_{{i},\mathbf{x}_w}}={c_{{i},\mathbf{x}_f}}$ to estimate their
values, which is commonly used in the boundary scheme of the LB
method~\cite{Guo3}. As a result, we ultimately derive the
distributions of the order parameters at the ghost node,
\begin{eqnarray}
{c_{{1},\mathbf{x}_f+\mathbf{e}_k\delta_t}}={c_{{1},\mathbf{x}_f}}+\frac{4\delta_x}{D}(\cos{\theta_{13}}{{c}_{1,\mathbf{x}_f}}{{c}_{3,\mathbf{x}_f}}
+\cos{\theta_{12}}{{c}_{1,\mathbf{x}_f}}{{c}_{2,\mathbf{x}_f}}), \nonumber\\
{c_{{2},\mathbf{x}_f+\mathbf{e}_k\delta_t}}={c_{{2},\mathbf{x}_f}}+\frac{4\delta_x}{D}(-\cos{\theta_{12}}{{c}_{1,\mathbf{x}_f}}{{c}_{2,\mathbf{x}_f}}
+\cos{\theta_{23}}{{c}_{2,\mathbf{x}_f}}{{c}_{3,\mathbf{x}_f}}), \nonumber\\
{c_{{3},\mathbf{x}_f+\mathbf{e}_k\delta_t}}={c_{{3},\mathbf{x}_f}}+\frac{4\delta_x}{D}(-\cos{\theta_{23}}{{c}_{2,\mathbf{x}_f}}{{c}_{3,\mathbf{x}_f}}
-\cos{\theta_{13}}{{c}_{1,\mathbf{x}_f}}{{c}_{3,\mathbf{x}_f}}).
\end{eqnarray}
From Eq. (42), the values of the order parameters at the ghost node
has been determined, and then $\nabla ^2 c_i$ at the fluid nodes
neighboring to solid wall can be computed by Eq. (20b).

In addition to the computation of $\nabla ^2 c_i$, the space
gradients $\nabla c_i$, $\nabla {\rho}$ and $\nabla ^2 \mu_i$ at the
fluid node $\mathbf{x}_f$ should also be given in the LB algorithm.
The evaluation of these gradients using Eqs. (20a) and (20b) require
the unknown information at the ghost node
$(\mathbf{x}_f+\mathbf{e}_k\delta_t)$, which can be determined based
on the symmetric rule with respect to the solid wall~\cite{Liang2},
\begin{eqnarray}
c_i(\mathbf{x}_f+\mathbf{e}_k\delta_t)=c_i(\mathbf{x}_f),\nonumber\\
\rho(\mathbf{x}_f+\mathbf{e}_k\delta_t)=\rho(\mathbf{x}_f),\nonumber\\
\mu_i(\mathbf{x}_f+\mathbf{e}_k\delta_t)=\mu_i(\mathbf{x}_f).
\end{eqnarray}
The scheme in Eq. (43) used here satisfies no flux condition, and
also can avoid unphysical mass and momentum transfer through the
solid boundary. The boundary conditions for the distribution
functions should also be specified in the implementation of the LB
method. In this work, we apply the half-way bounce back boundary
scheme for dealing with the solid wall, which is realized by setting
the unknown distribution functions to be the ones in the opposite
directions~\cite{Liang2}
\begin{equation}
\begin{split}
 &
 f^i_{\bar{k}}(\textbf{x}_f, t+\delta_t)=f'^i_k(\textbf{x}_f, t),\\&
 g_{\bar{k}}(\textbf{x}_f, t+\delta_t)=g'_k(\textbf{x}_f, t),
\end{split}
\end{equation}
where ${\bar{k}}$ is the opposite direction of ${{k}}$, $f'^i_k$ and
$g'_k$ are the postcollision distribution functions given by
\begin{equation}
\begin{split}
& {f'^i_k}({\textbf{x}_f}, t) = {f^i_k}(\textbf{x}_f,t)-
\frac{1}{\tau_i}\left[
{{f^i_k}(\textbf{x}_f,t)- f_k^{i,eq}(\textbf{x}_f,t)} \right] + {\delta_t}{F^i_k}(\textbf{x}_f,t),\\
& {g'_k}(\textbf{x}_f,t) = {g}_k(\textbf{x}_f,t)
-\frac{1}{\tau_g}[{{g}_k(\textbf{x}_f,t) -
g_k^{eq}(\textbf{x}_f,t)}]+ {\delta _t}G_k(\textbf{x}_f,t).
\end{split}
\end{equation}
The boundary scheme has been proven to preserve the second-order
numerical accuracy in the space, which retains the same accuracy as
that of the LB method.

\begin{figure}
\centering
\includegraphics[width=3.2in,height=1.4in]{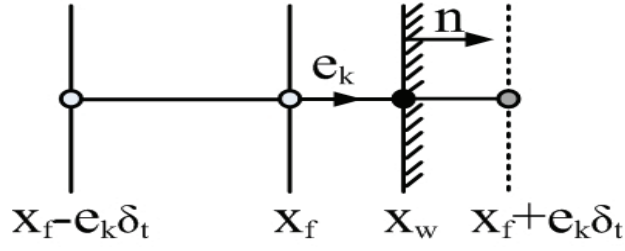}
 \tiny\caption{Schematic of lattice node and solid wall boundary.}
\end{figure}
\section{Numerical Results and discussions}\label{sec:Results}

In this section, we first perform the simulations of some basic
three-phase flow problems to validate the proposed LB model coupled
with the wetting boundary condition. These typical problems involve
partially wettable solid surfaces, which include the spreading of
binary drops, the spreading of a compound drop, and the shear of a
compound liquid drop. Here we also conduct a detailed comparison
between the present numerical results with the analytical solutions
or some available results. As an application, at last we use the
present method to study the dynamics of a compound drop impact on a
solid circular cylinder.

\begin{figure}
\centering
\includegraphics[width=5.5in,height=1.47in]{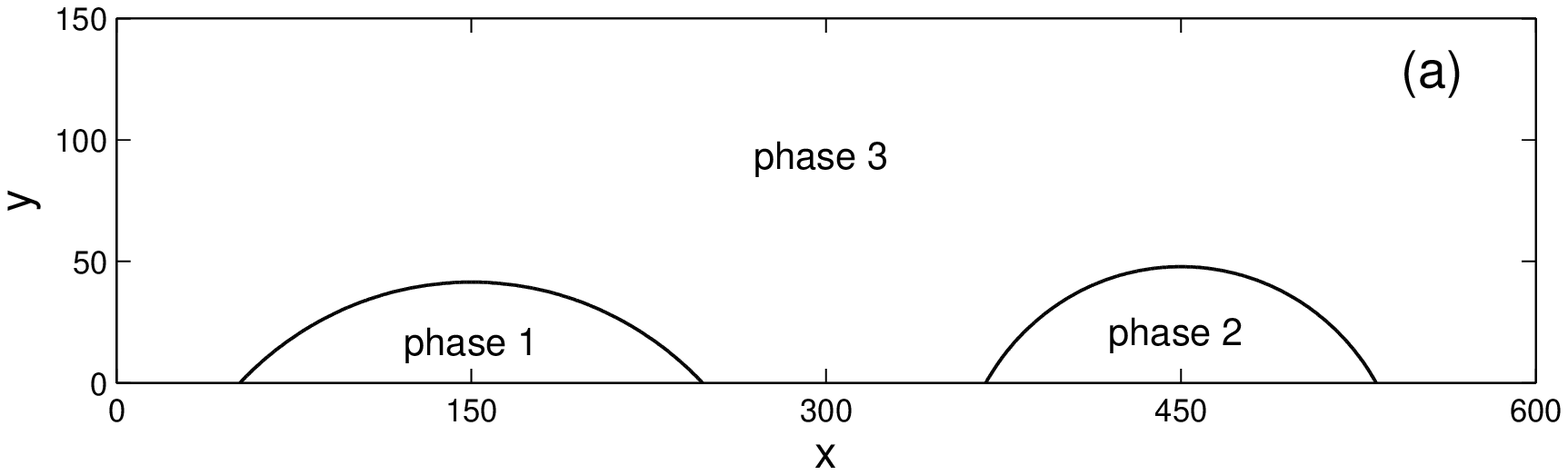}\\
\includegraphics[width=5.5in,height=1.47in]{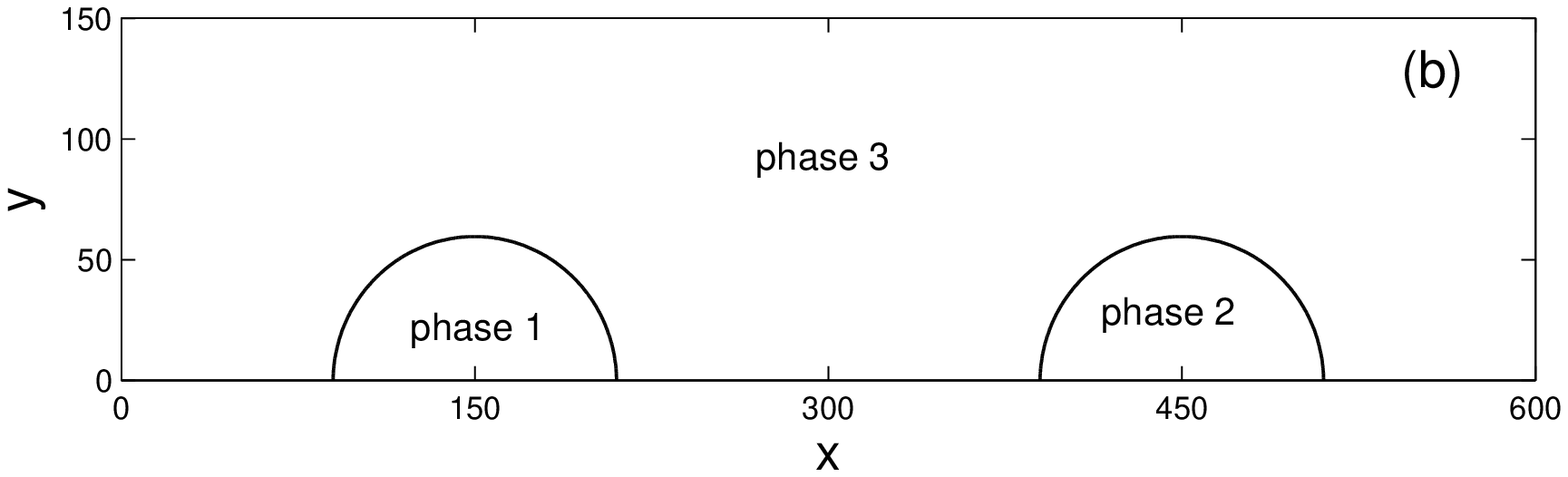}\\
\includegraphics[width=5.5in,height=1.47in]{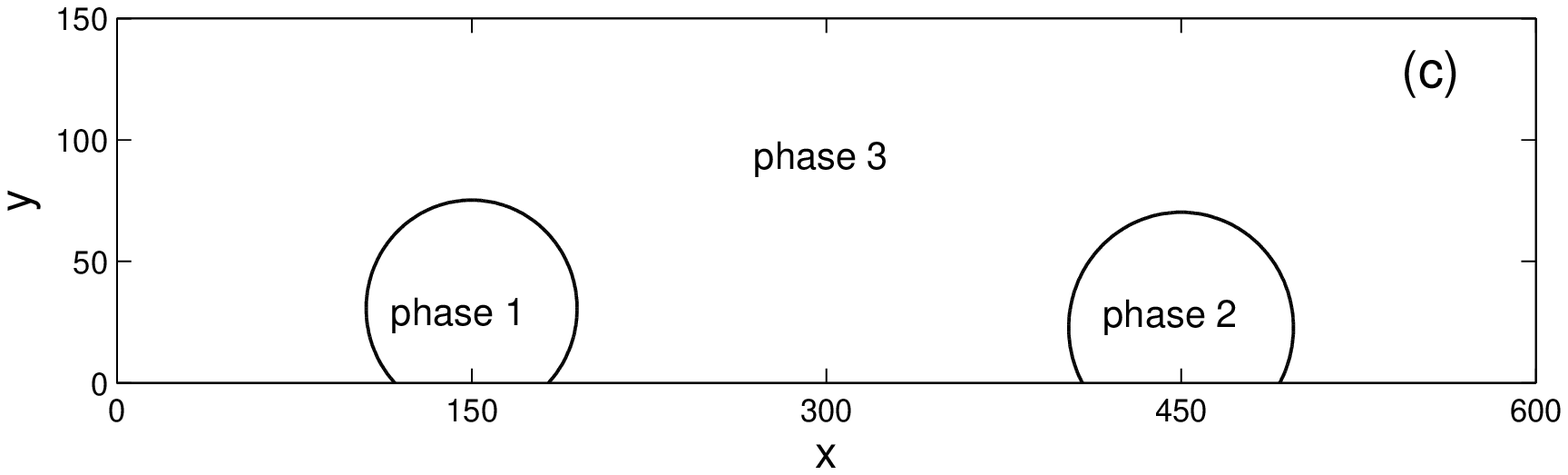}
 \tiny\caption{The equilibrium configurations of binary drops on the horizontal substrate with various contact angles,
 (a) $(\theta_{13}, \theta_{23})=(45^\circ, 60^\circ)$; (b) $(\theta_{13}, \theta_{23}) =(90^\circ, 90^\circ)$; (c) $(\theta_{13}, \theta_{23})=(135^\circ, 120^\circ)$.}
\end{figure}

\begin{figure}
\centering
\includegraphics[width=3.2in,height=2.8in]{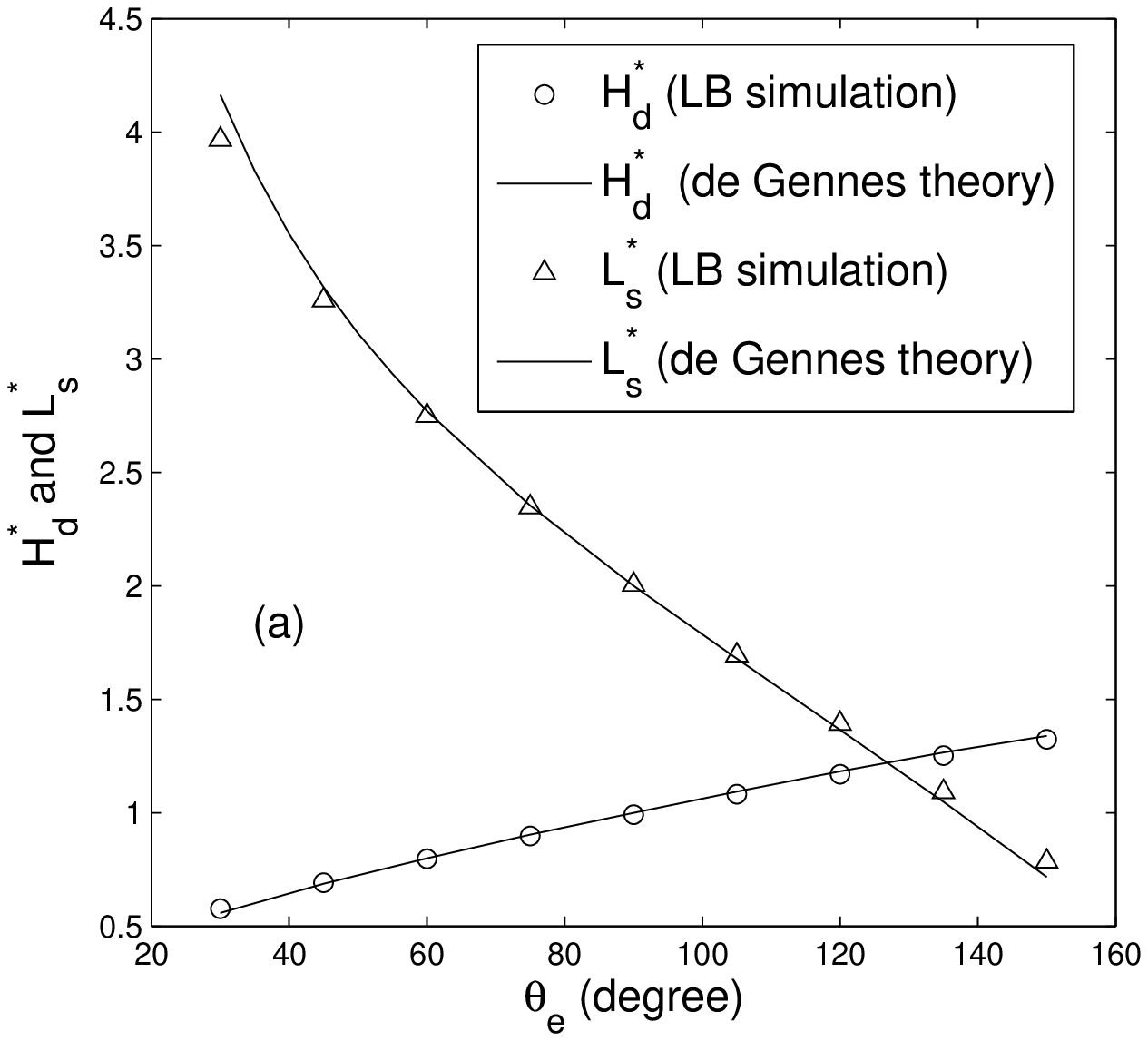}
\includegraphics[width=3.2in,height=2.8in]{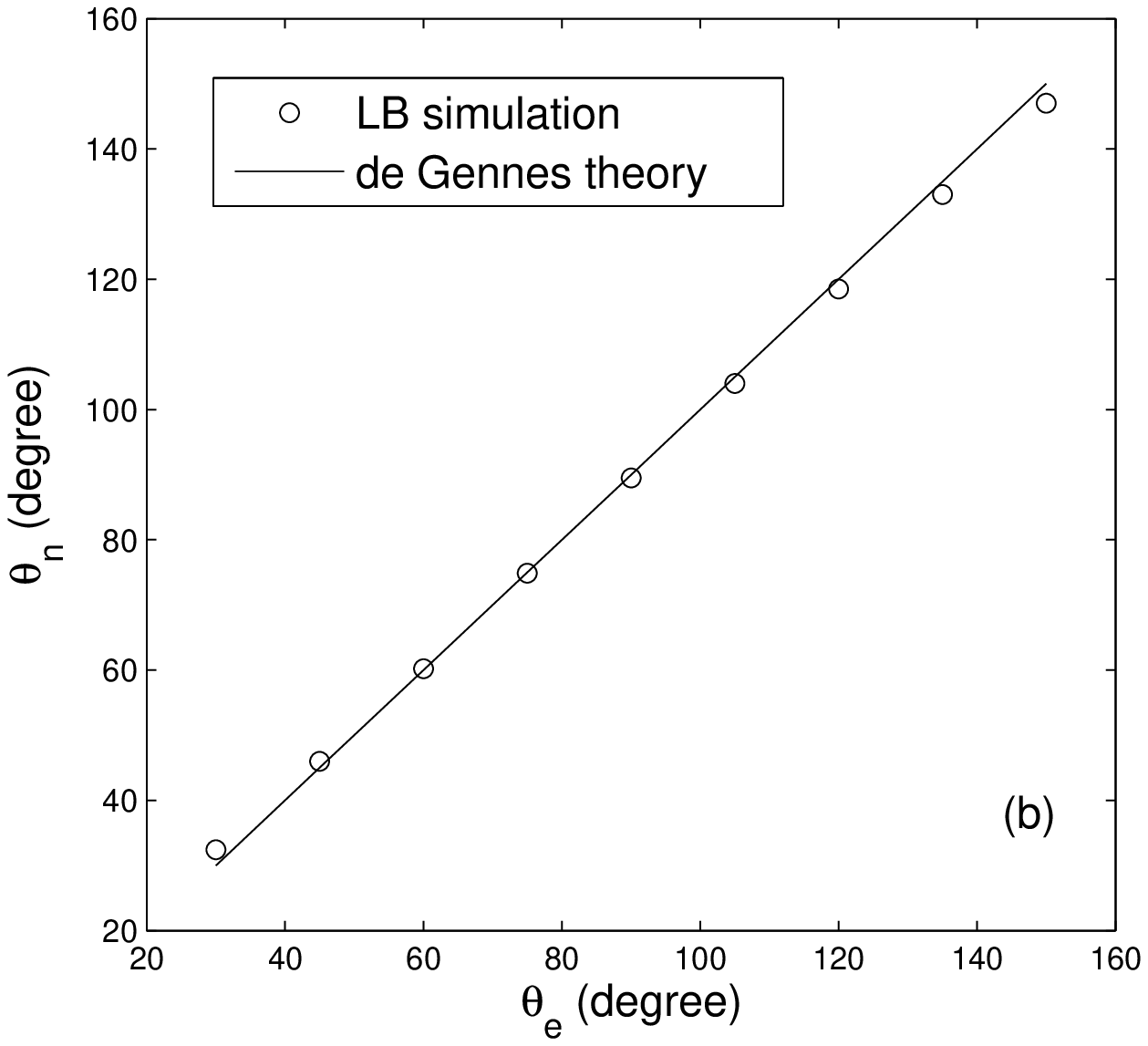}\\
 \tiny\caption{(a) comparison of the equilibrium drop height and spreading length as a function of contact angle between LB simulation and the de Gennes theory~\cite{Gennes};
 (b) the measured contact angle ($\theta_n$) versus the theoretical value. The
height $H_d^*$ and length $L_s^*$ has been normalized by the
characteristic scale $R$.}
\end{figure}

\subsection{The spreading of binary drops}

 We first consider a simple case of the spreading of two liquid drops with different densities on the horizontal
 substrate, which is a fundamental three-phase flow problem to validate the numerical method~\cite{Dong}. The initial gap between
 binary drops is assumed to be sufficiently large, such that the interaction between them is very weak, and then can be neglected.
 In this case, each drop has the equilibrium pattern familiar from the one in two-phase case, which
allows us to quantitatively compare the present numerical results
with the de Gennes theory~\cite{Gennes}. The physical system
considered here is a rectangular domain with a size of $L\times W$,
where $L$ and $W$ are the length and width of the domain, and
$L/W=4$. Initially, two liquid drops with radius $R=0.1L$ are placed
on a partially wetting substrate, and their centers are respectively
located at $(x_1, y_1)=(0.25L, 0)$ and $(x_2, y_2)=(0.75L, 0)$. In
our simulations, the length $L$ is set to be 600 lattice unit, and
the order parameters are initialized by
\begin{equation}
\begin{split}
 c_1(x,y)&=0.5+0.5\tanh \frac{2\left[R-(x-x_1)^2-(y-y_1)^2\right]}{D},\\
 c_2(x,y)&= 0.5+0.5\tanh \frac{2\left[R-(x-x_2)^2-(y-y_2)^2\right]}{D},
 \end{split}
\end{equation}
which makes their values to be smooth across the interface. The
densities of binary drops are 10 and 5, with an ambient fluid having
the density of 1. Some other physical parameters in the simulation
are fixed as $D=4$, $\sigma_{12}=\sigma_{13}=\sigma_{23}=0.1$,
$\tau_1=\tau_2=\tau_g=0.8$, and $M_0=0.01$. The periodic boundary
condition is used in the horizontal direction, and we impose the
half-way bounce back boundary condition for the bottom and top wall.
The wetting boundary condition is also applied at the bottom
boundary. When the system is released, it begins to evolve, and
eventually reaches its equilibrium state. We mainly focus on the
equilibrium configurations of two liquid drops and the contact angle
effect.

\begin{figure}
\centering
\includegraphics[width=5.5in,height=1.47in]{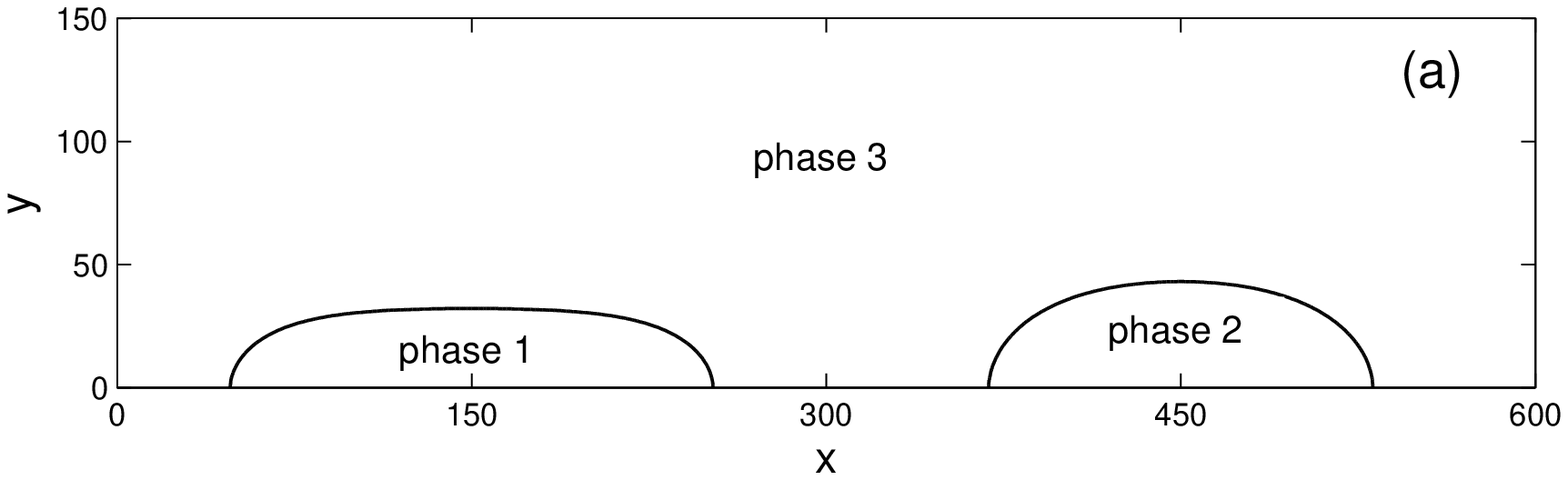}
\includegraphics[width=5.5in,height=1.47in]{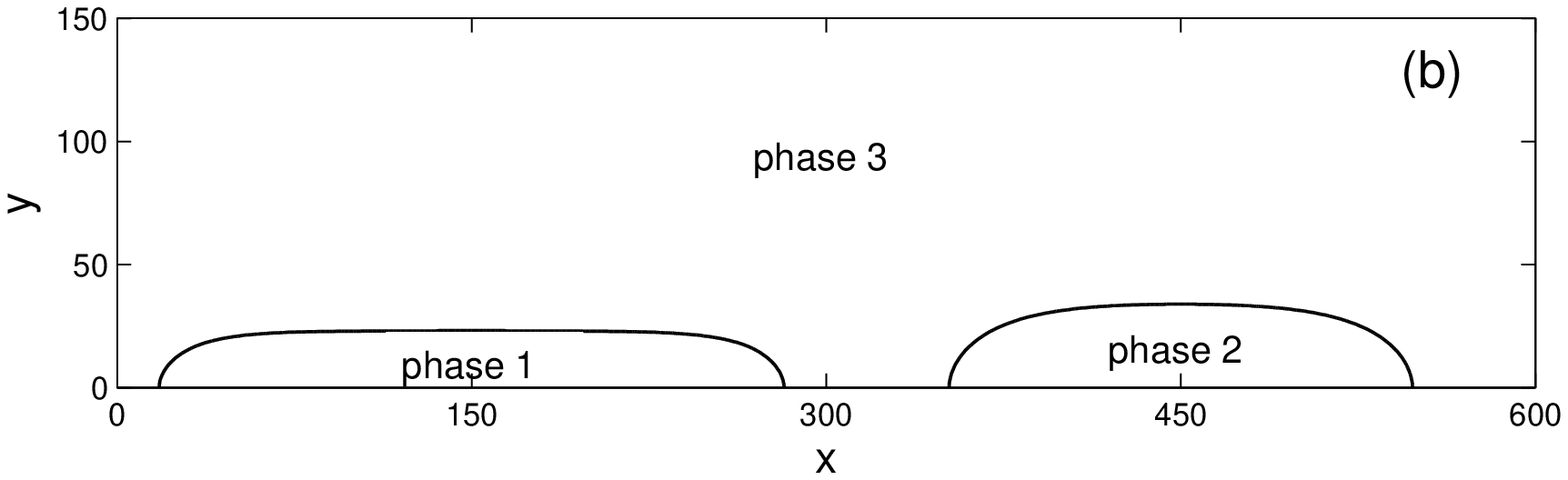}\\
 \tiny\caption{The equilibrium shapes of binary drops at different gravity forces, (a) $g=2\times10^{-5}$; (b) $g=4\times10^{-5}$.}
\end{figure}

Figure 2 depicts the equilibrium shapes of binary liquid drops on
the horizontal substrate with three typical groups of contact angles
$(\theta_{13}, \theta_{23})=(45^\circ, 60^\circ)$, $(\theta_{13},
\theta_{23}) =(90^\circ, 90^\circ)$, and $(\theta_{13},
\theta_{23})=(135^\circ, 120^\circ)$. Note that when $\theta_{13}$
and $\theta_{23}$ have been given, the value of $\theta_{12}$ can be
determined from the constraint (39), and here the interface of each
drop is marked by the contour levels $c_i=0.5$. From Fig. 2, we can
observe that each drop intends to adhere the wall, forming a
circular cap when the contact angle is less than $90^\circ$. And it
raises on the wall for the contact angle larger than $90^\circ$,
while it almost keeps resting on the wall when the contact angle is
$90^\circ$. The behavior of the drop is in line with the
expectation~\cite{Gennes}. In addition, we can see that the
equilibrium shape of each drop in the three-phase system is
qualitatively consistent with the one that drop alone exists in an
ambient fluid. To give a quantitative comparison, we also measure
the spreading length ($L_s$) between the fluid and substrate and the
drop height ($H_d$) at the equilibrium state, as illustrated in Fig.
3. According to the de Gennes theory~\cite{Gennes}, the equilibrium
spreading length and height can be determined by the following
relations,
\begin{equation}
 L_s=2R\sqrt{\frac{\pi/2}{\theta_e-\sin \theta_e \cos \theta_e}}\sin
 \theta_e,~~
 H_d=R\sqrt{\frac{\pi/2}{\theta_e-\sin \theta_e \cos
 \theta_e}}(1-\cos \theta_e),
\end{equation}
where $\theta_e$ is the equilibrium contact angle. We plotted the
dimensionless equilibrium spreading length and height as a function
of the static contact angle in Fig. 3(a), where $L_s^*$ and $H_d^*$
have been normalized by the characteristic length $R$. For a
comparison, the theoretical results from Eq. (47) are also
presented. It is shown that the numerical results are in good
agreement with the corresponding analytical solutions. We further
measured the numerical contact angle ($\theta_n$) based on the
geometrical relation $\theta_n=2\arctan 2H_d/L_s$ and presented the
results in Fig. 3(b). It is found that the numerical predictions of
the contact angles are consistent with the theoretical values.

We now consider the influence of gravity on the equilibrium profile
of the above three-phase system. Due to the existing of the gravity,
the shape of each drop greatly depends on the relative importance of
three force including the gravity force, the surface tension force
between fluids, and the adhering force between fluid and solid. For
the convenience of discussion, one can introduce a particular
length, which is also named as the capillary length denoted by
$\gamma^{-1}$. The capillary length is estimated by comparing
relative magnitudes of the Laplace pressure and the hydrostatic
pressure, and then can be defined by
$\gamma^{-1}=\sqrt{\frac{\sigma}{\rho_l g}}$~\cite{Gennes}, where
$\sigma$ is the surface tension between two fluids, $\rho_l$ is the
density of the liquid fluid, and $g$ is the gravitational
acceleration. When the drop radius is much smaller than
$\gamma^{-1}$, the surface tension force is dominant over the
gravity, and then the drop takes on a shape of a circular cap at
equilibrium. While when the drop size sufficiently exceeds
$\gamma^{-1}$, the gravity force is a crucial force that comes into
play in the system, which then results in the formation of a puddle
at the equilibrium state~\cite{Gennes}. Based on the de Gennes
theory~\cite{Gennes}, the asymptotic thickness ($H$) of a puddle can
be analytically expressed as
\begin{equation}
H=2\gamma^{-1}\sin \frac{\theta_e}{2}.
\end{equation}
We have simulated the spreading of binary drops the horizontal
substrate in the presence of the gravity force. In our simulation,
the gravity force $\mathbf{G}=(0,-\rho g)$ is applied to all three
fluids, and the contact angles ${\theta_{13}}$ and $\theta_{23}$ are
assumed to be $90^\circ$. The other physical parameters are set as
those in the previous situations. Figure 4 shows the equilibrium
shapes of binary drops at two different gravitational accelerations.
The result with the case of zero gravity is presented in Fig. 2(b).
As expected, the drop forms a circular shape at zero gravity, and it
intends to spread on the substrate when the gravity is imposed. The
shape of the drop becomes flatted with the increase of the gravity,
and a puddle can be formed when the gravity is sufficiently large.
We also quantitatively measured the asymptotic thickness of each
drop from its equilibrium profile, and presented the results with
different gravity values in Fig. 5. For a comparison, the
corresponding theoretical results given in Eq. (48) are also
presented. It can be found from Fig. 5 that our numerical results
agree well with the theoretical solutions at large gravity values,
while they diverge from the theoretical solutions at small gravity
values. These obvious discrepancies can be attributed to the fact
that theoretical thickness curve, computed based on Eq. (48) is only
valid when the gravity is dominant~\cite{Dong, Gennes}.

\begin{figure}
\centering
\includegraphics[width=3.5in,height=3.0in]{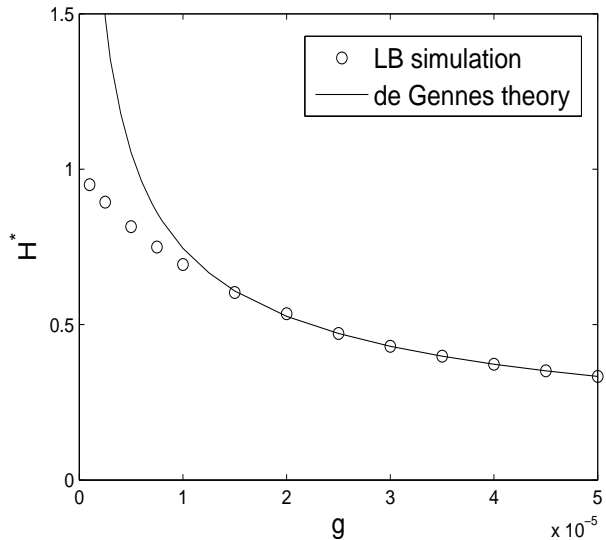}
 \tiny\caption{Comparison of the asymptotic thickness as a function
of gravity between the LB simulation and the de Gennes
theory~\cite{Gennes}. The thickness $H^*$ has been normalized by the
characteristic length $R$. }
\end{figure}

\subsection{The spreading of a compound drop}
\begin{figure}
\centering
\includegraphics[width=5.0in,height=2.4in]{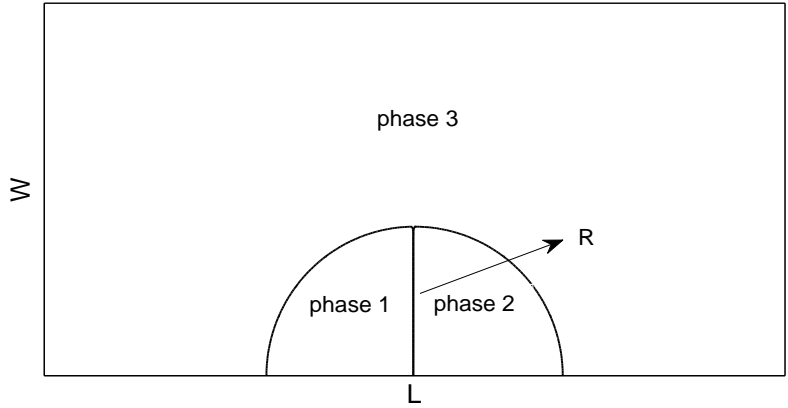}
 \tiny\caption{The initial setup of the spreading of a compound drop. }
\end{figure}
The interaction between two phases in the previous situation is very
small. In this subsection, we will consider a case of the spreading
of a compound drop on the solid substrate to validate the wetting
conditions for ternary fluids, where the interactions among three
fluids are very strong, and the equilibrium profile of the system
significantly depends on the contact angles among fluids and
solid~\cite{Said,Ding2}. The initial setup of the physical problem
is shown in Fig. 6, in which $L$ and $W$ are the length and width of
a rectangle domain, the 1-th and 2-th phase fluids constitute a
semi-circle compound drop with the radius $R$ surrounded by the 3-th
phase. The boundary conditions are adopted as those of the previous
simulation. Some physical parameters in this test are given as
$L=2W=300$, $R=60$, and the remaining ones used are set as those of
the previous case. Here we focus on the equilibrium shape in a wide
range of the contact angles $\theta_{13}$ and $\theta_{23}$ from
$30^\circ$ to $150^\circ$, and the value of the contact angle
$\theta_{12}$ can be determined from the relation (39). We first
investigated the effect of the contact angle $\theta_{12}$ with a
fixed contact angle $\theta_{23}=90^\circ$. Figure 7 depicts the
equilibrium configurations of a compound liquid drop at various
contact angles $\theta_{13}$. We can observe that its equilibrium
shape is remarkably affected by the contact angle $\theta_{13}$. For
the situation of $\theta_{13}=60^\circ$, the 1-th phase intends to
spread on the wall, and it also partially moves underneath the 2-th
phase, due to the value of $\theta_{13}$ smaller than $90^\circ$.
With the increase of the $\theta_{13}$, the 1-th phase begins to
shrink on the substrate and the region occupied by the phase 1 and
the solid is also reduced. As a result, it plumps up with respect to
the 2-th phase at equilibrium and the extent increases with the
$\theta_{13}$, as can be clearly seen in Figs. 7(b) and (c). When
the $\theta_{13}$ is sufficiently large, the 1-th phase intends to
migrate away from the solid at equilibrium~\cite{Dong}. In the
present numerical experiment, we indeed observe this phenomenon
shown in Fig. 7(d), where the 1-th phase moves above the 2-th phase
and is no longer in contact with the solid. We also numerically
measured the contact angles $\theta_{12}$, $\theta_{13}$ and
$\theta_{23}$ between two fluids of the system and the solid, and it
is found that all the numerical results are consistent with the
initially prescribed values. In addition, we further measured the
equilibrium three-phase contact angles $\varphi_1$, $\varphi_2$,
$\varphi_3$ at the triple junction, as illustrated in Fig. 7(a). The
values of $\varphi_1$, $\varphi_2$, $\varphi_3$ depend on the
relative importance of three surface tensions between two fluids of
a three-phase system~\cite{Kim},
\begin{equation}
\frac{\sin{\varphi_1}}{\sigma_{23}}=\frac{\sin{\varphi_2}}{\sigma_{13}}=\frac{\sin{\varphi_3}}{\sigma_{12}}
\end{equation}
and also obviously satisfy the relation
$\varphi_1+\varphi_2+\varphi_3=2\pi$. Here the surface tension ratio
is fixed to be 1, therefore the analytical solutions for the
$\varphi_1$, $\varphi_2$, $\varphi_3$ are all equal to $120^\circ$.
We also have computed the $\varphi_1$, $\varphi_2$, $\varphi_3$ from
the equilibrium profile corresponding to each value of the
$\theta_{13}$, and find that all the numerical predictions of the
contact angles $\varphi_i$ $(i=1,2,3)$ approximate to $120^\circ$,
which have good agreement with the analytical solutions.

\begin{figure}
\centering
\includegraphics[width=2.8in,height=1.36in]{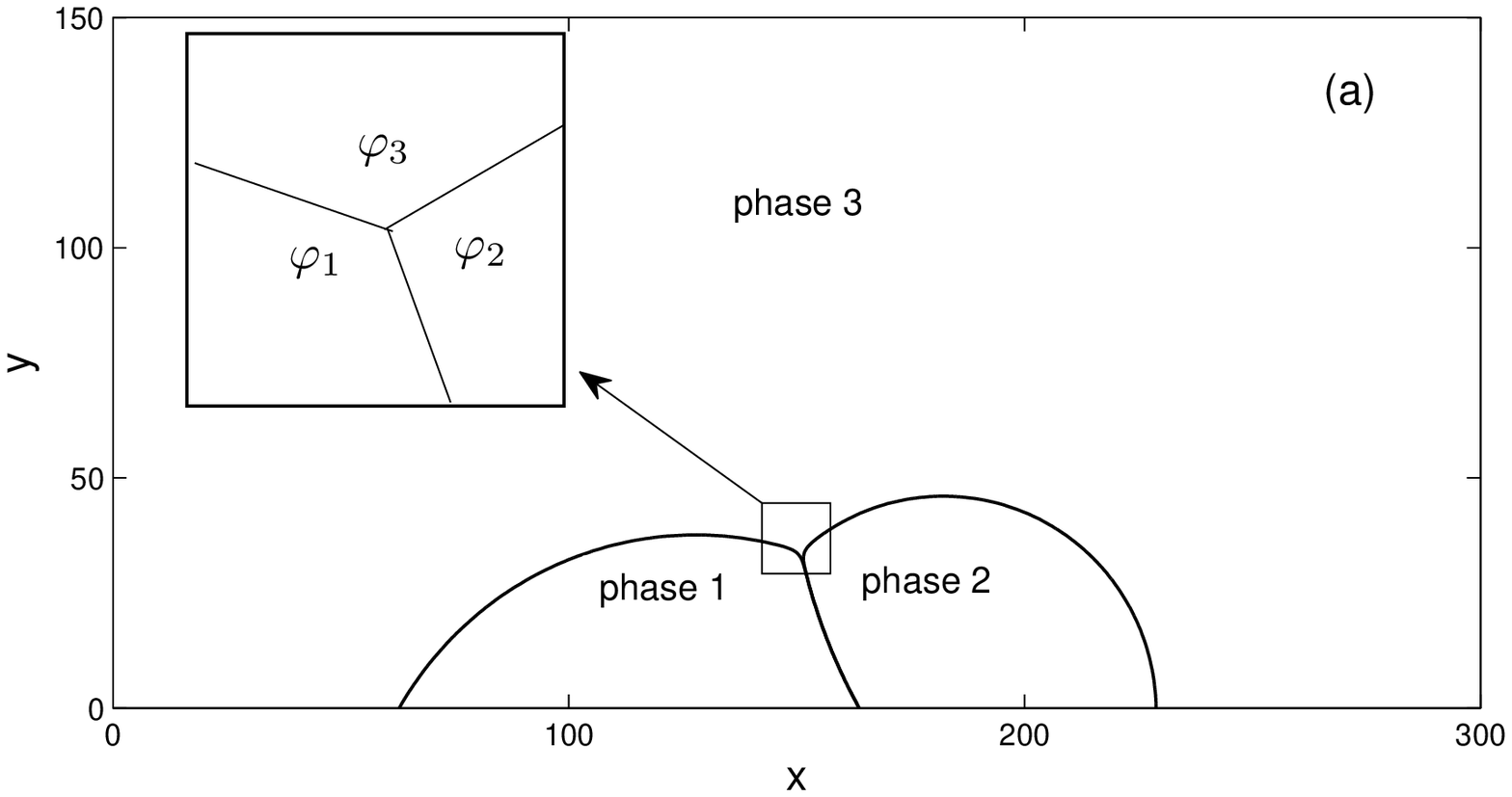}
\includegraphics[width=2.8in,height=1.36in]{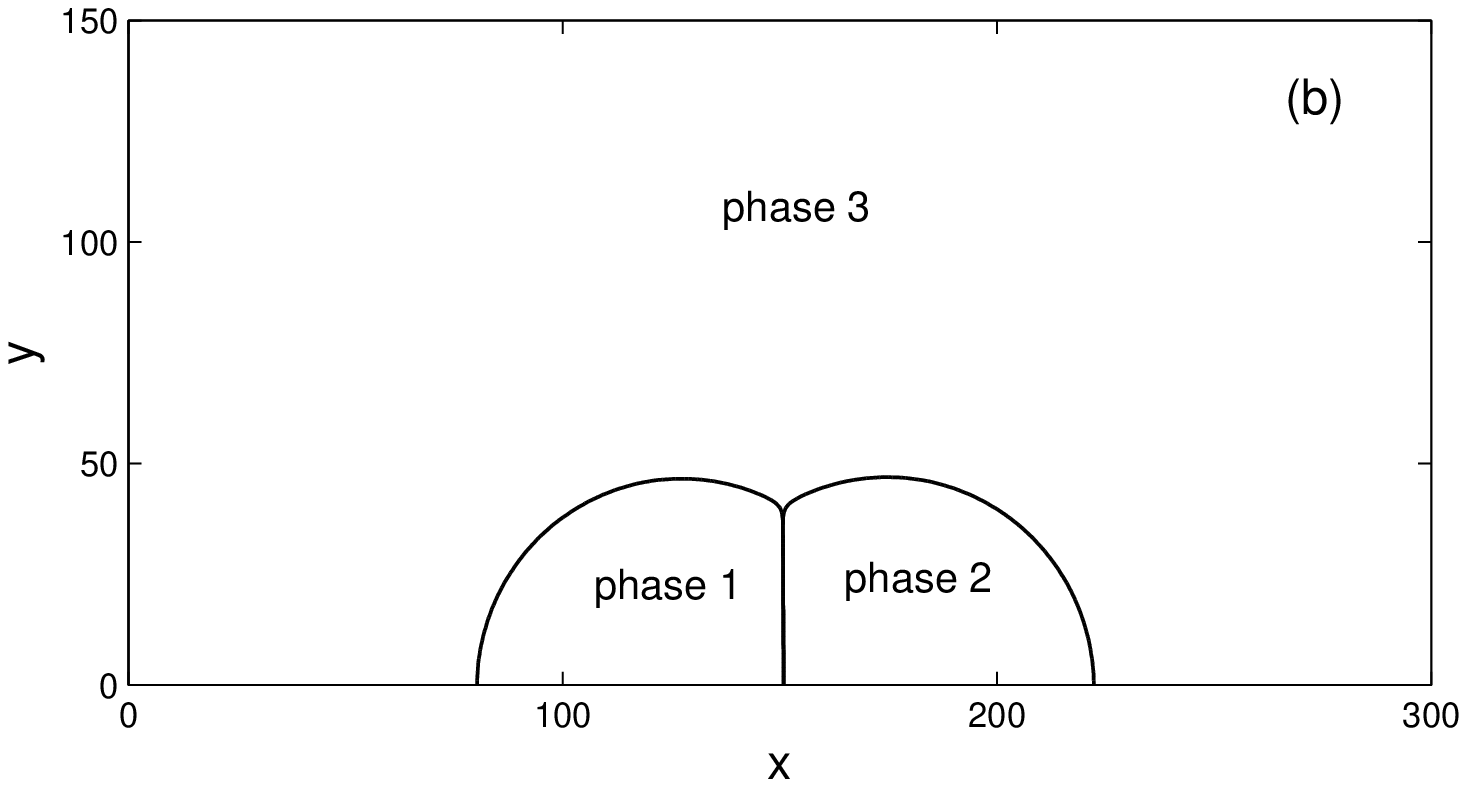}\\
\includegraphics[width=2.8in,height=1.36in]{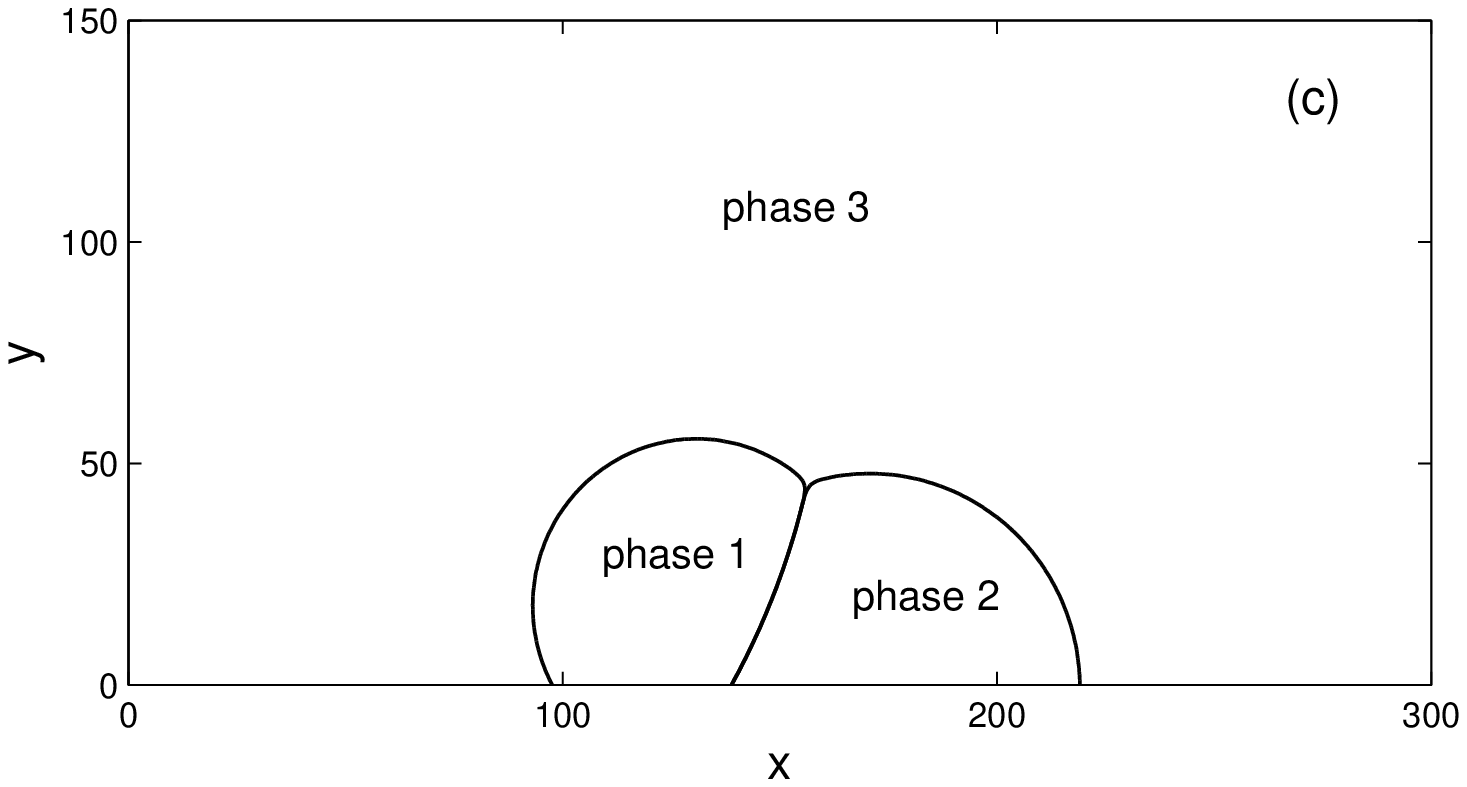}
\includegraphics[width=2.8in,height=1.36in]{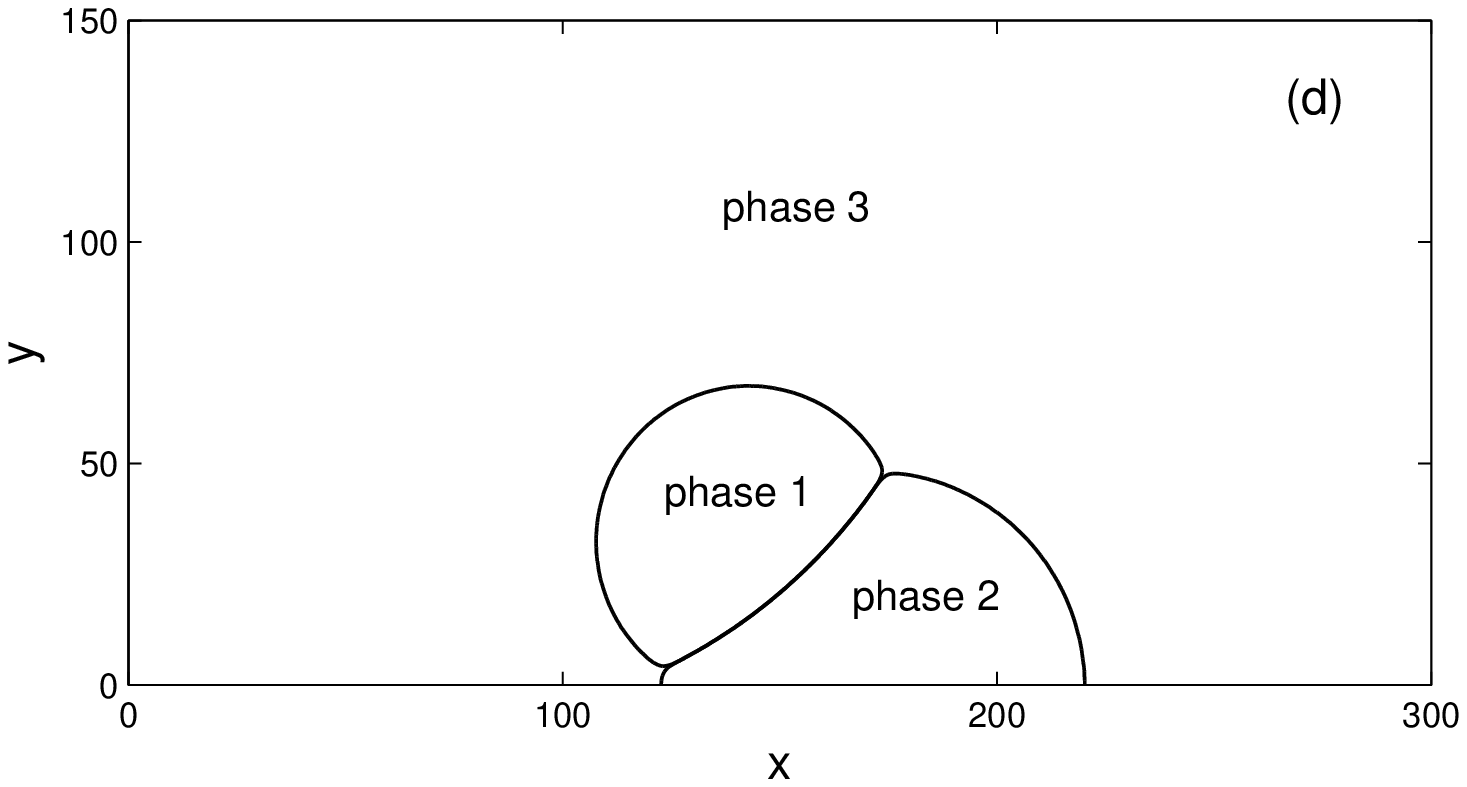}
 \tiny\caption{The equilibrium configurations of a compound liquid drop at a fixed contact angle of $\theta_{23}=90^\circ$ and various contact angles $\theta_{13}$, (a) $\theta_{13}=60^\circ$;
  (b) $\theta_{13}=90^\circ$; (c) $\theta_{13}=120^\circ$; (d) $\theta_{13}=150^\circ$.}
\end{figure}

\begin{figure}
\centering
\includegraphics[width=2.8in,height=1.35in]{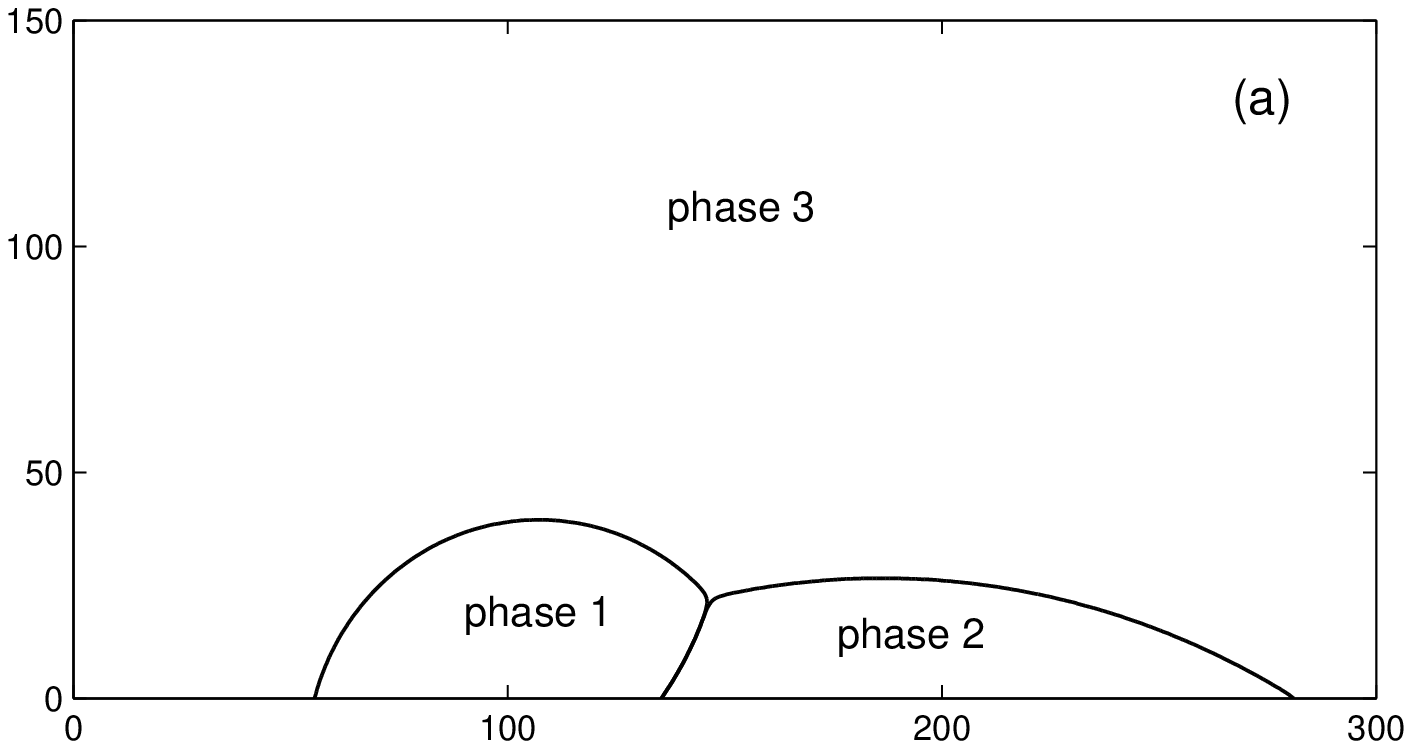}
\includegraphics[width=2.8in,height=1.35in]{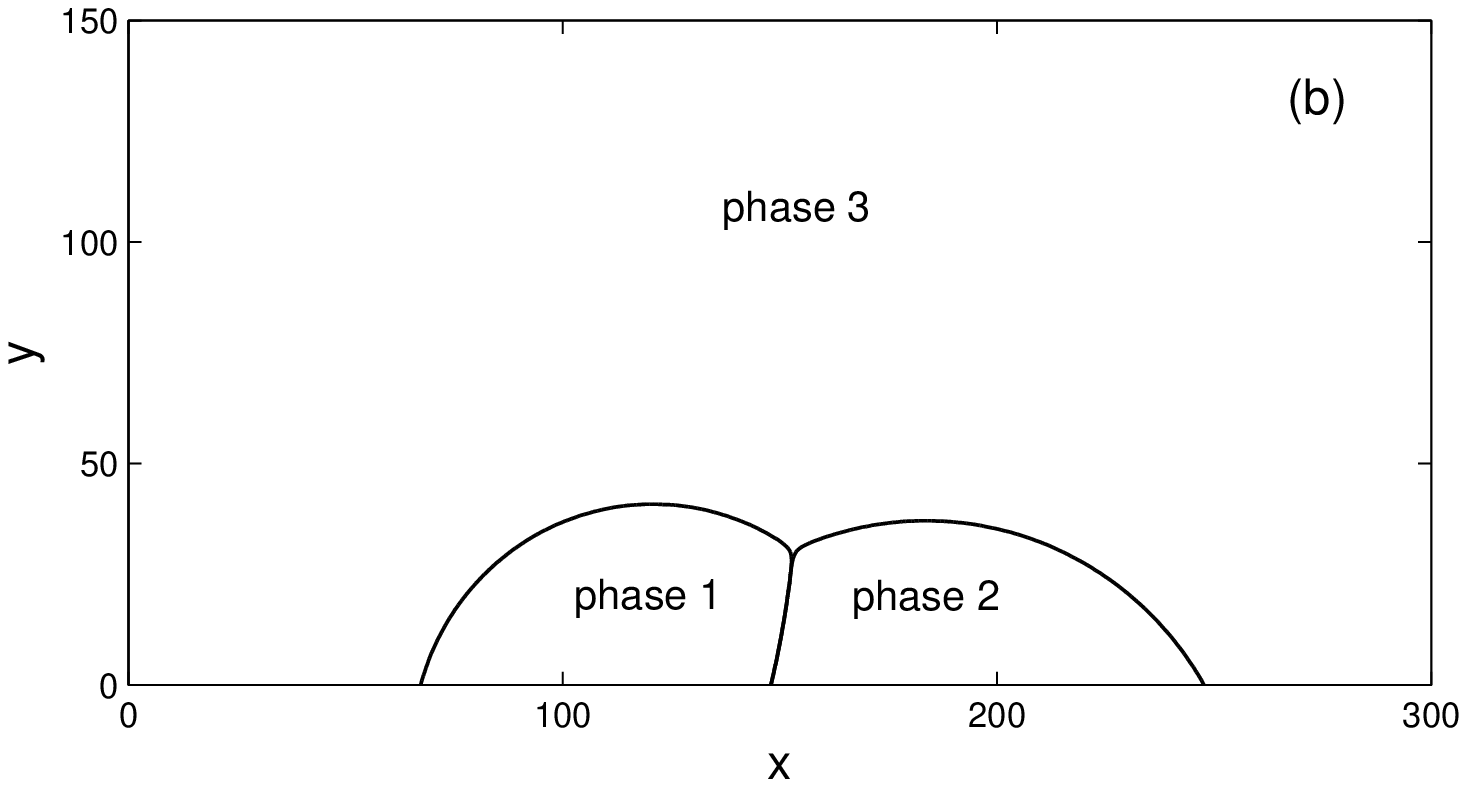}\\
\includegraphics[width=2.8in,height=1.35in]{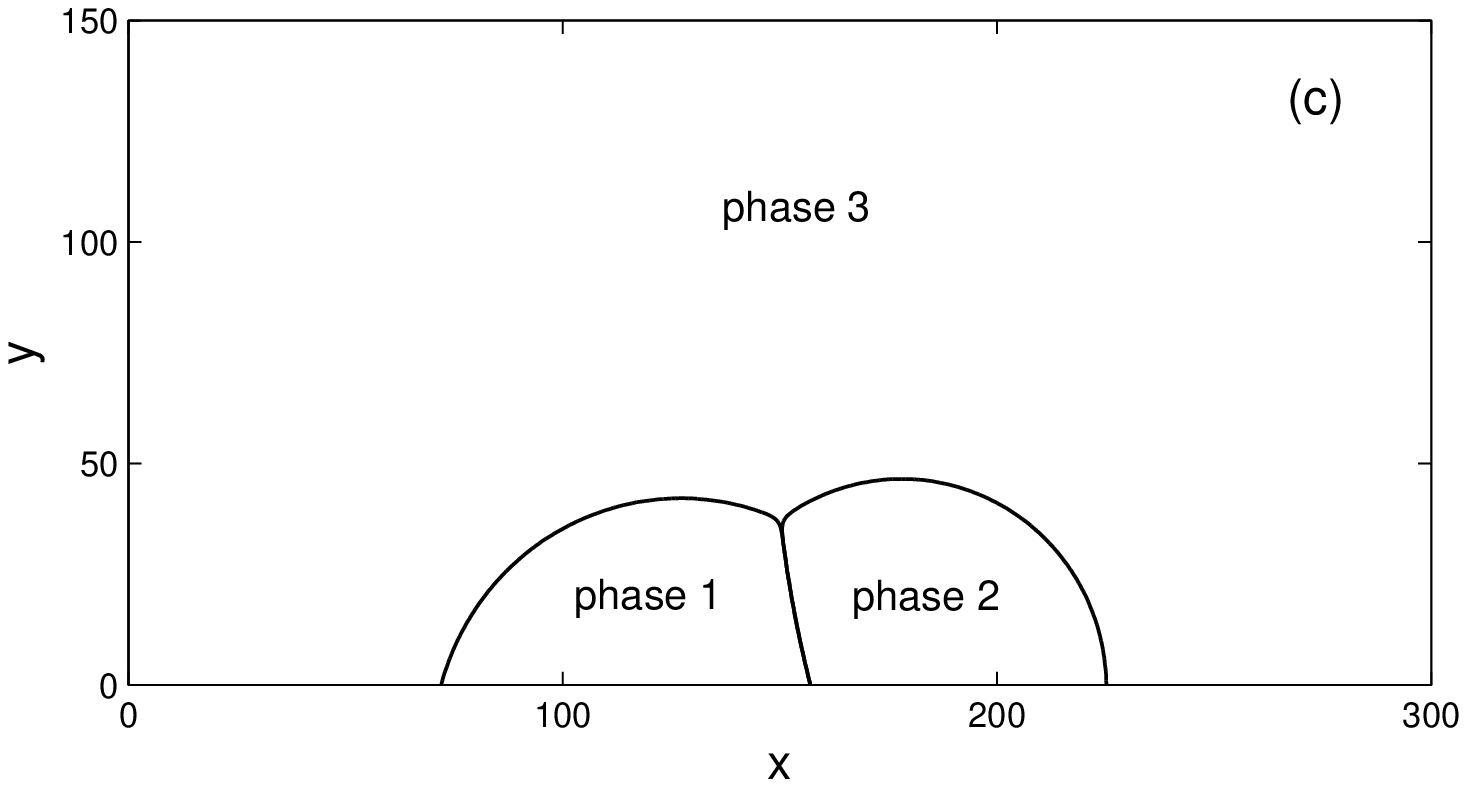}
\includegraphics[width=2.8in,height=1.35in]{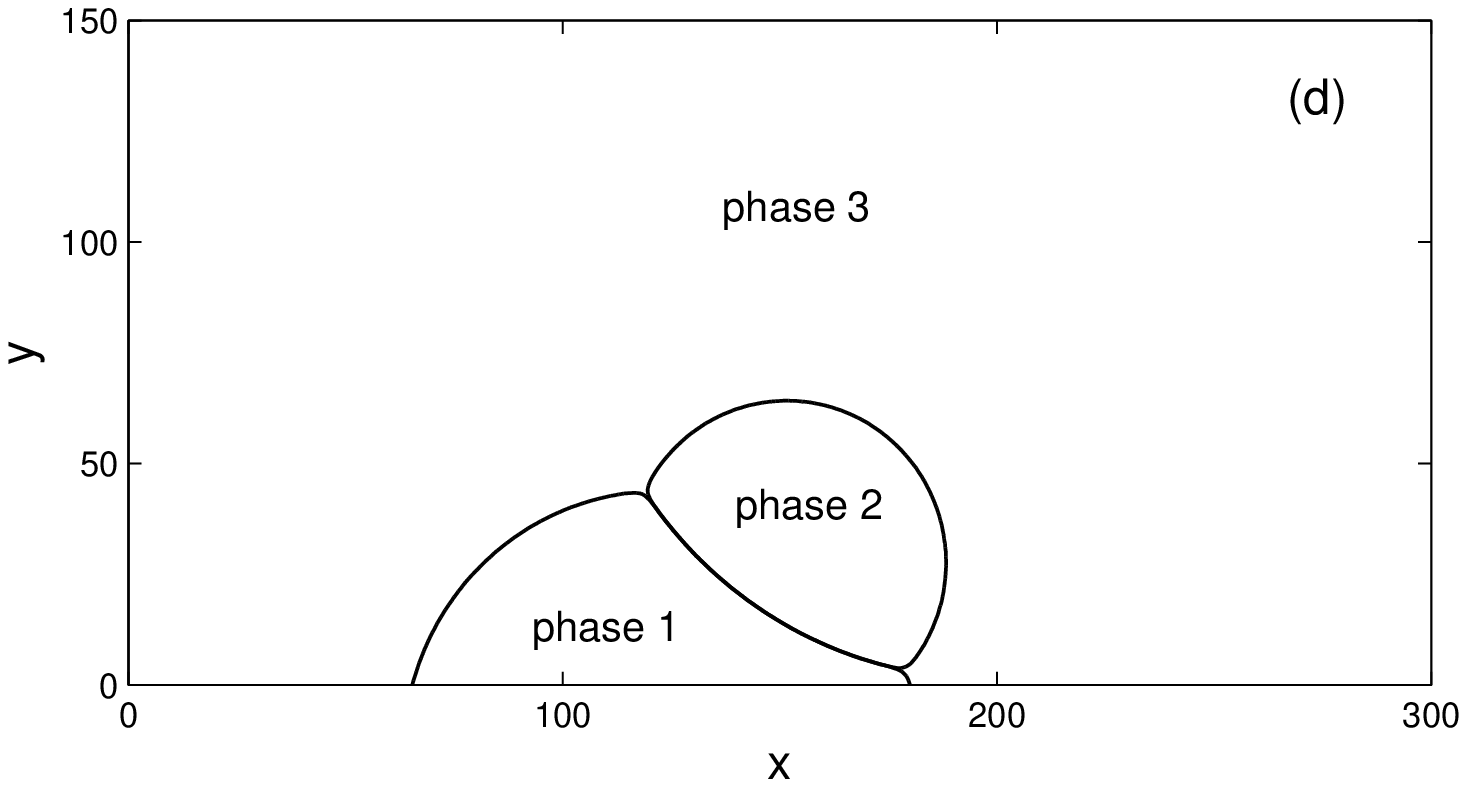}
 \tiny\caption{The equilibrium configuration of a compound liquid drop at a fixed contact angle of $\theta_{13}=75^\circ$ and various contact angles $\theta_{23}$,
 (a) $\theta_{23}=30^\circ$; (b) $\theta_{23}=60^\circ$; (c) $\theta_{23}=90^\circ$; (d) $\theta_{23}=135^\circ$.}
\end{figure}

We now focus on the effect of the contact angle $\theta_{23}$ with a
fixed value of $\theta_{13}=75^\circ$. Figure 8 shows the
equilibrium configurations of a compound liquid drop at four typical
contact angles $\theta_{23}$. It can be observed that the contact
angle $\theta_{23}$ dramatically influences the equilibrium shape of
a compound drop. As the $\theta_{23}$ increases, the spreading of
the 2-th phase on the wall is reduced, and it becomes more and more
plump with respect to the neighbouring phase 1 at the equilibrium.
Particularly, when the $\theta_{23}$ is large enough, the 2-th phase
would depart from the solid wall and is located on the upside of the
1-th phase fluid. We also quantitatively measured the contact angles
$\theta_{ij}$ $(1\leq i<j\leq3)$ and $\varphi_{i}$ $(1\leq i\leq3)$,
and it is found that the numerical values of the $\theta_{ij}$ are
in accordance with the initially given ones, and the values of the
$\varphi_{i}$ also conform to Eq. (49).

\begin{figure}
\centering
\includegraphics[width=2.8in,height=1.2in]{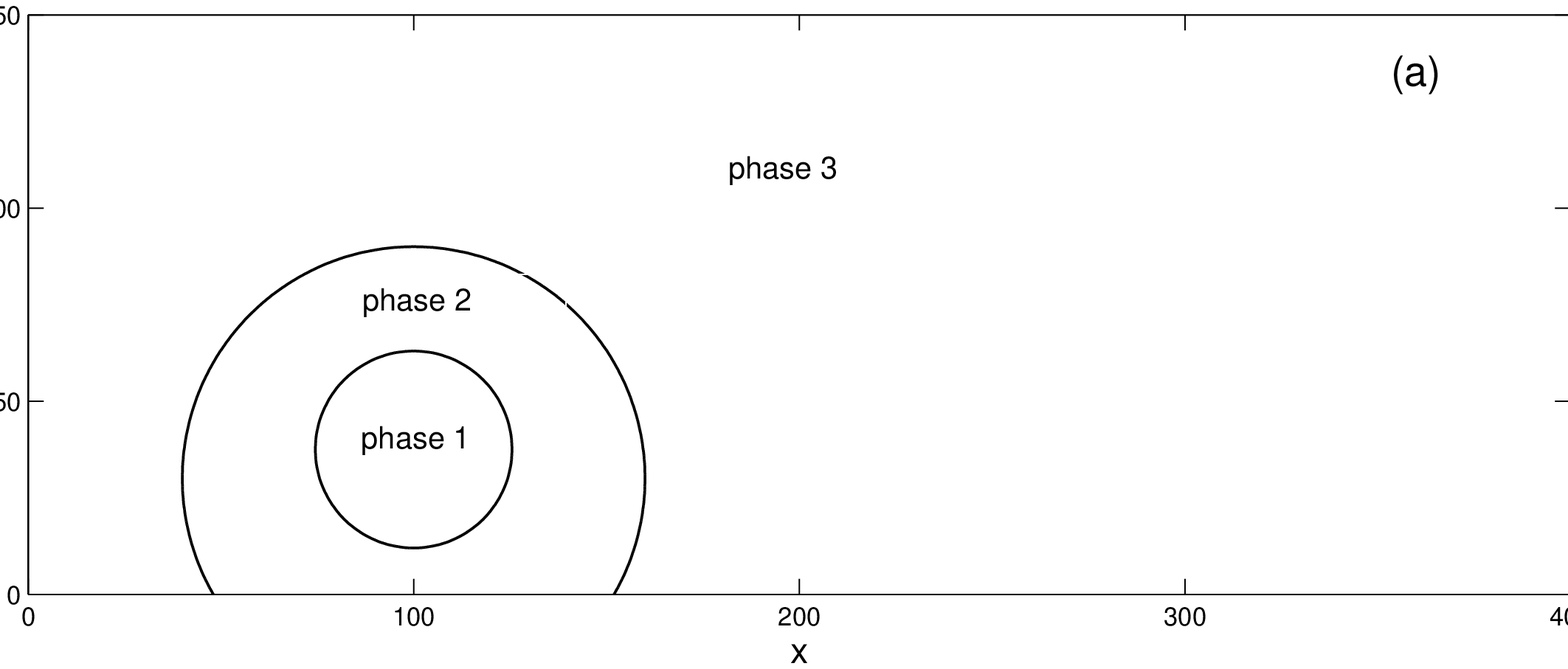}~~~~~~~~~~~
\includegraphics[width=2.8in,height=1.2in]{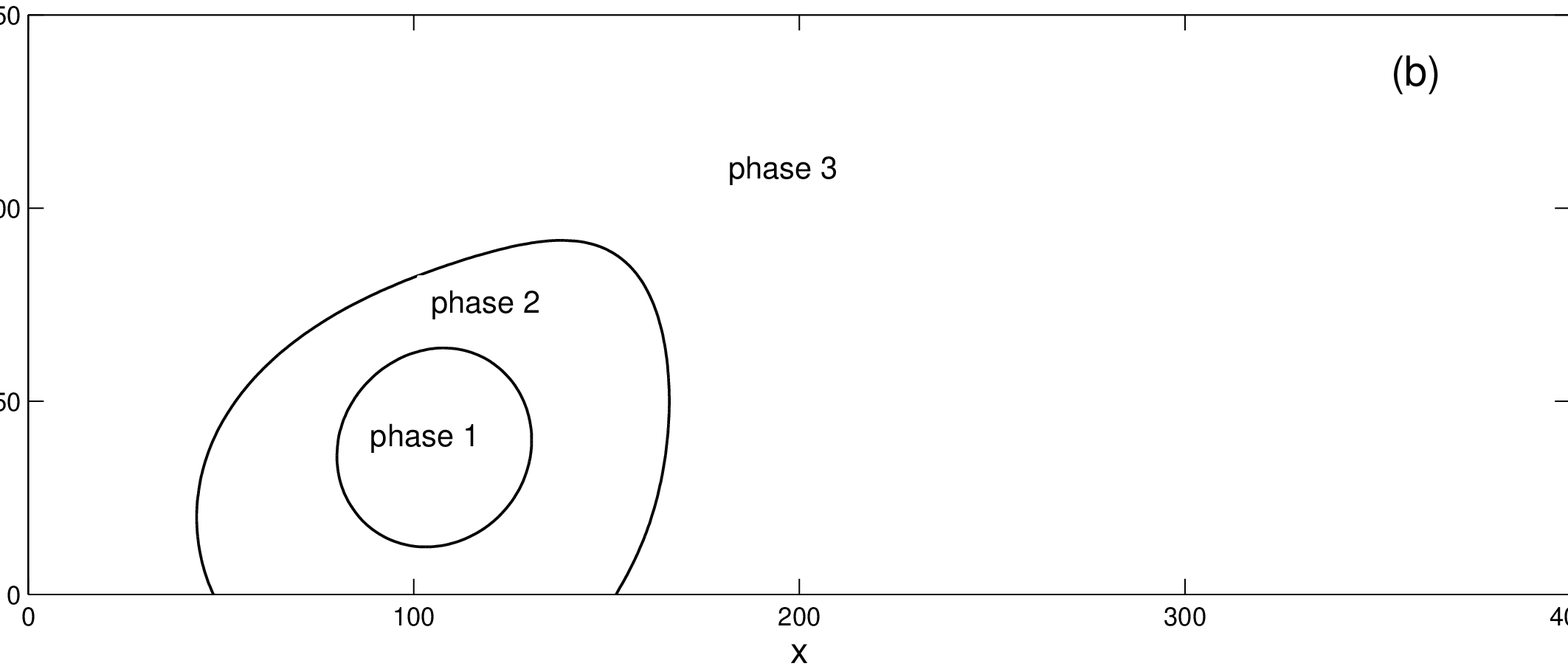}
\includegraphics[width=2.8in,height=1.2in]{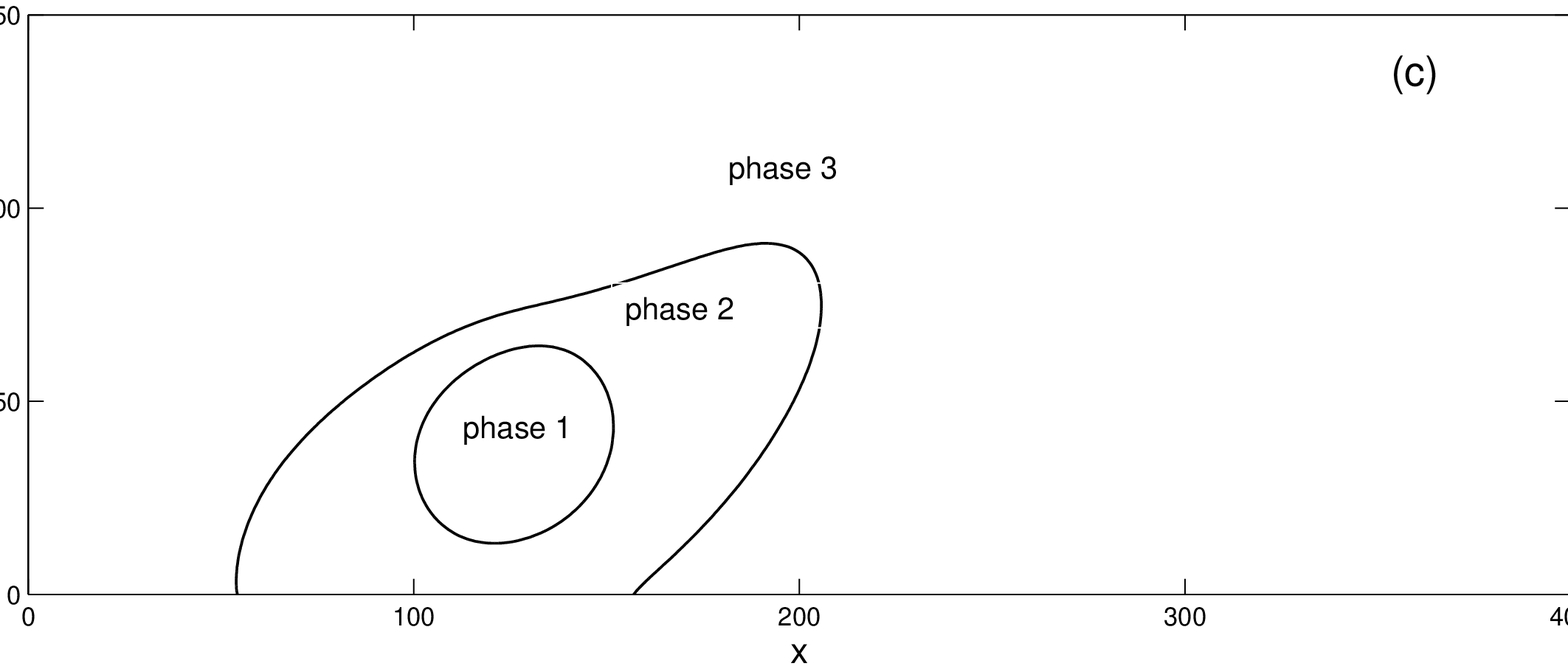}~~~~~~~~~~~
\includegraphics[width=2.8in,height=1.2in]{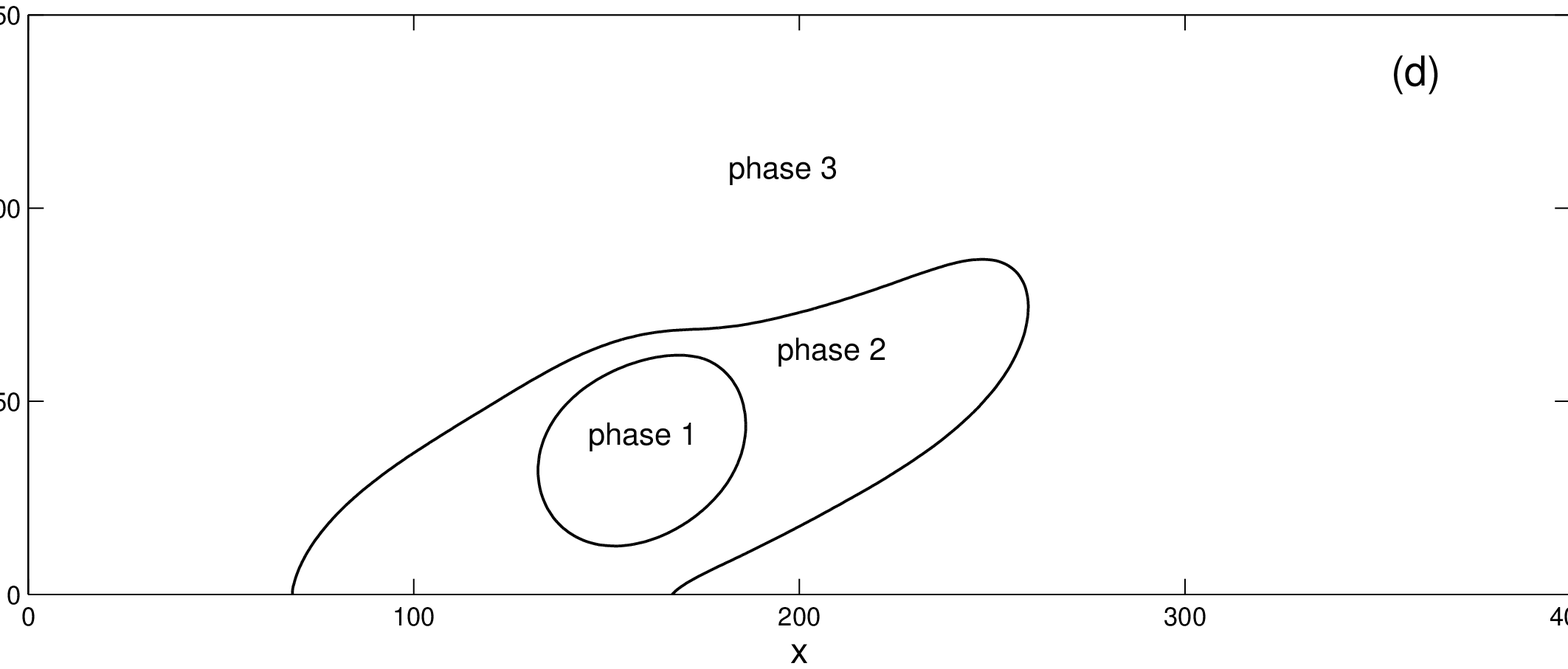}
\includegraphics[width=2.8in,height=1.2in]{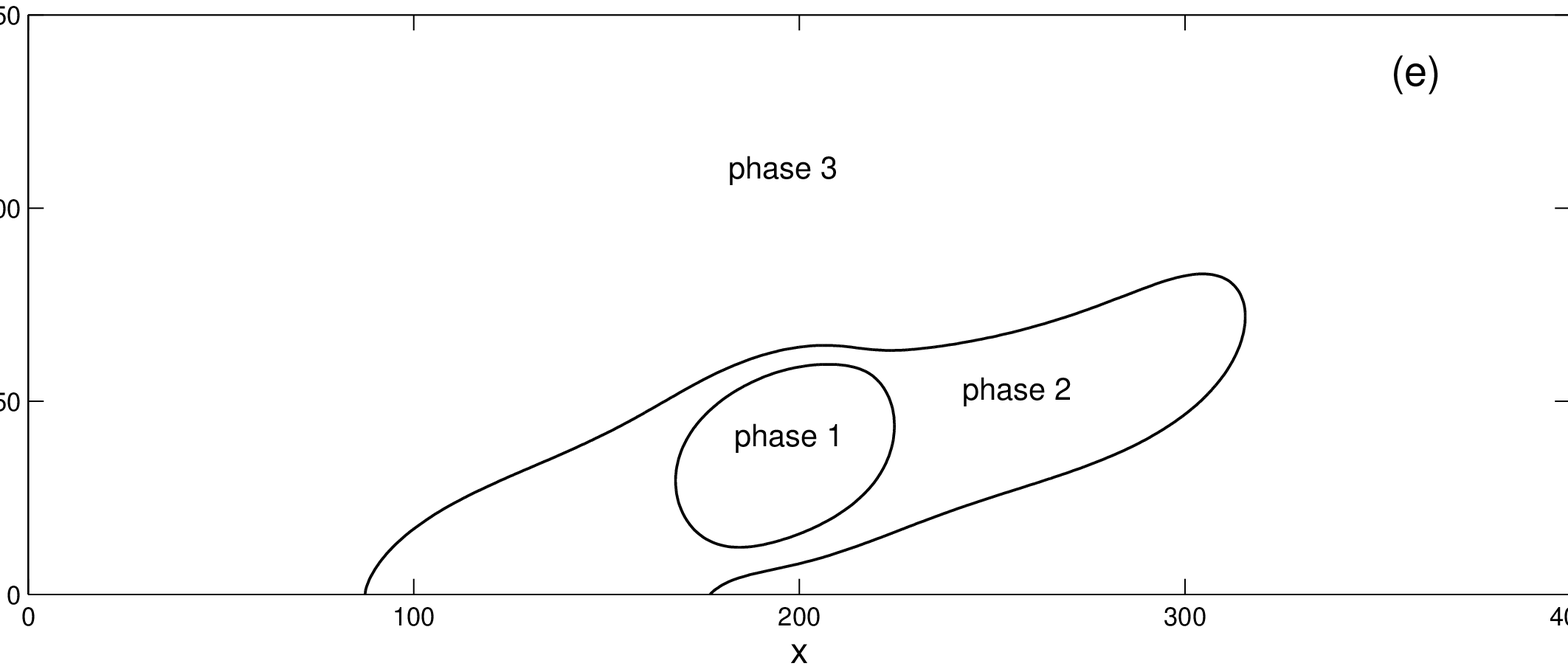}~~~~~~~~~~~
\includegraphics[width=2.8in,height=1.2in]{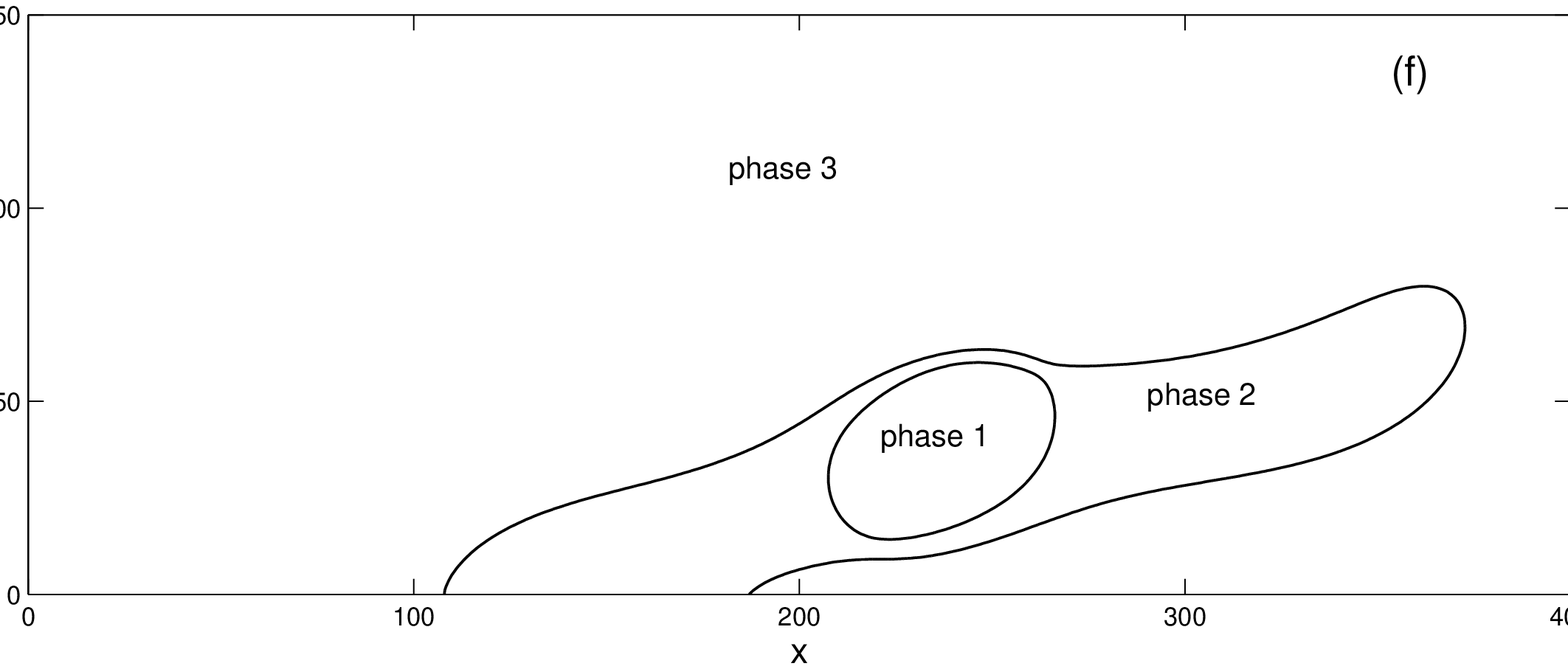}
 \tiny\caption{The evolution of interface patterns of a compound liquid drop,(a) t=0; (b) t=10000; (c) t=15000; (d) t=20000; (e) t=25000; (f) t=30000.}
\end{figure}

\subsection{The shear of a compound liquid drop}

The shear of a compound drop has extensive applications in the
fields of biomedical models of leukocyte~\cite{Kan} and
oil-water-gas displacement process~\cite{Schleizer}, and due to the
rareness of a suitable numerical approach, very few numerical
studies on this subject can be avaiable. In this subsection, we will
simulate a compound liquid drop adhering to the wall subject to the
shear flow by the three-phase LB model coupled with the present
wetting boundary scheme. Initially, a liquid compound drop rests on
the substrate of a rectangular domain with the grid $N_x\times
N_y=400\times150$, and a constant horizontal velocity $U_0$ with the
value of 0.1 is applied at the upper boundary. The profiles of the
order parameters can be initialized by
\begin{equation}
\begin{split}
 c_1(x,y)&=0.5+0.5\tanh \frac{2\left[R_1-(x-x_{c1})^2+(y-y_{c1})^2\right]}{D},\\
 c_2(x,y)&=0.5+0.5\tanh \frac{2\left[R_2-(x-x_{c2})^2+(y-y_{c2})^2\right]}{D}-c_1(x,y),
 \end{split}
\end{equation}
where $R_1$ is the radius of the circular drop for the 1-th phase
with a value of $25.5$, $(x_{c1},y_{c1})=(37.5,100)$ is its central
coordinate, $R_2=60$ is the radius of the 2-th phase drop,
$(x_{c2},y_{c2})=(30,100)$, and the interface thickness $D$ is 4.
The other physical parameters in our simulations are set to be
$\rho_1=\rho_2=\rho_3=1$~\cite{Yshi},
$\sigma_{12}=\sigma_{13}=\sigma_{23}=0.01$,
$\tau_1=\tau_2=\tau_g=0.8$, and $M_0=0.001$. The problem is periodic
in the $x$-direction with the periodic boundary conditions applied
for the left and right boundaries. The lower boundary is the solid
wall imposed by the no-slip bounce back boundary condition. In
addition, to describe the wettability property of the wall material,
the present wetting boundary scheme is also applied at the lower
boundary. Here three contact angles considered are given as
$\theta_{12}=30^\circ$, $\theta_{13}=68.53^\circ$, and
$\theta_{23}=120^\circ$. As for the upper boundary, it is velocity
boundary which is treat by the nonequilibrium extrapolation scheme
proposed by Guo {\it{ et. al.}}~\cite{Guo3, Guo4}. The
nonequilibrium extrapolation scheme has been successfully applied in
single-phase flows~\cite{Guo1}, and also shows good performance in
the study of the two-phase flows~\cite{Liang2, Gong}. Here we extend
this scheme directly to the three-phase flow situation. The main
idea of the scheme is that the distribution function at the boundary
is divided into the equilibrium part at the local boundary node and
the nonequilibrium part at the neighbouring fluid node~\cite{Guo3}.
Based on this idea~\cite{Guo3}, the upper boundary condition for
three-phase flows can be realized by
\begin{equation}
\begin{split}
{f^i_k}({\mathbf{x}_u},t)&=f_k^{i,eq}({\bf{x}}_u,t)+f_k^{i,neq}({\bf{x}}_f,t),~~i=1,~2,\\
{g_k}({\mathbf{x}_u},t)&=g_k^{eq}({\bf{x}}_u,t)+g_k^{neq}({\bf{x}}_f,t),
 \end{split}
\end{equation}
where ${\mathbf{x}_u}$ is the node at the upper boundary,
${\mathbf{x}_f}$ is its neighbouring fluid node, the nonequilibrium
parts $f_k^{i,neq}({\bf{x}}_f,t)$ and $g_k^{neq}({\bf{x}}_f,t)$ can
be given by
\begin{equation}
\begin{split}
f_k^{i,neq}({\bf{x}}_f,t)&={{f^i_k}(\textbf{x}_f,t)-
f_k^{i,eq}(\textbf{x}_f,t)},~~i=1,~2,\\
g_k^{neq}({\bf{x}}_f,t)&={{g_k}(\textbf{x}_f,t)-
g_k^{eq}(\textbf{x}_f,t)},
\end{split}
\end{equation}
that have been known, and the equilibrium parts
$f_k^{i,eq}({\bf{x}}_u,t)$ and $g_k^{eq}({\bf{x}}_u,t)$ can be
expressed as
\begin{subequations}
\begin{equation}
f_{k}^{i,eq}({\bf{x}}_u,t) =\left\{
\begin{array}{ll}
 c_i({\bf{x}}_u,t)  + ({\omega_k} - 1)\eta {\mu_i}({\bf{x}}_u,t),                                       & \textrm{ $k=0$}   \\
 {\omega_k}\eta{\mu_i}({\bf{x}}_u,t)  + {\omega_k}c_i({\bf{x}}_u,t){{{\textbf{e}_k} \cdot {\bf{u}}}({\bf{x}}_u,t) \over {c_s^2}}, & \textrm{
 $k\neq0$},~~i=1,~2,
\end{array}
\right.
\end{equation}
\begin{equation}
g_k^{eq}({\bf{x}}_u,t)=\left\{
\begin{array}{ll}
{p({\bf{x}}_u,t) \over {c_s^2}}({\omega _k} - 1) + \rho({\bf{x}}_u,t){s_k}[{\bf{u}}({\bf{x}}_u,t)],              & \textrm{ $k=0$},    \\
{p({\bf{x}}_u,t) \over {c_s^2}}{\omega _k} + \rho({\bf{x}}_u,t){s_k}[{\bf{u}}({\bf{x}}_u,t)],                    & \textrm{ $k\neq0$}, \\
\end{array}
\right.
\end{equation}
\end{subequations}
where ${\bf{u}}({\bf{x}}_u,t)=(U_0,0)$, and the unknown macroscopic
quantities $c_i({\bf{x}}_u,t)$, $\mu({\bf{x}}_u,t)$,
$p({\bf{x}}_u,t)$, and $\rho({\bf{x}}_u,t)$ in Eqs. (53a) and (53b)
are determined by the interpolation
$\chi({\bf{x}}_u,t)=\chi({\bf{x}}_f,t)$, where $\chi$ represent the
above quantities. Figure 9 depicts the time evolution of interface
patterns of a compound liquid drop. It is found that the drop of the
2-th phase has a deformation under the shear force of its
surrounding fluid, and becomes more and more elongated with time
along the direction of the shear velocity. Obviously, the upper
portion of the drop moves much faster than its lower portion. Also,
it is found that the inner drop of the 1-th phase undergoes an
interfacial deformation, while the extent is obviously reduced due
to the small shear interaction. Our present numerical results are
qualitatively consistent with those of the previous
study~\cite{Yshi}.

\subsection{A compound drop impact on a solid circular cylinder}

\begin{figure}
\centering
\includegraphics[width=2.4in,height=2.75in]{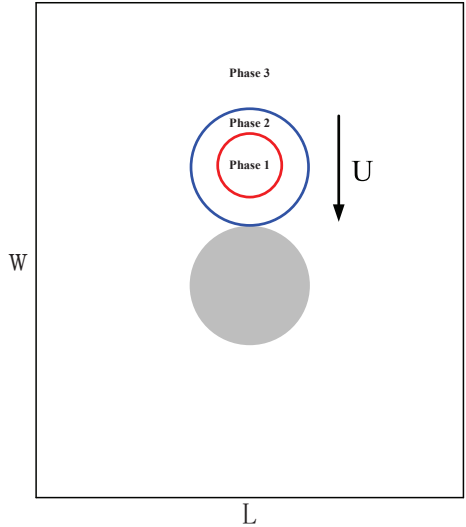}\\
 \tiny\caption{The schematic illustration for a compound drop impact on a solid circular cylinder. }
\end{figure}

At last, to show the versatility of the present wetting boundary
condition in dealing with three-phase fluid-solid systems, we
consider a complex problem of a compound drop impact on a solid
circular cylinder. Drop impact on a flat or curved substrate has
great relevance to many technical applications, such as oil
recovery, ink jet printing and and coating~\cite{Yarin, Lefebvre}.
Additionally, it can also serve as an important multiphase benckmark
problem, which involves very fascinating interfacial phenomena,
including spreading, splashing, bouncing, etc~\cite{Yarin}. Due to
its importance, drop impact on a solid target has been investigated
extensively using the experimental and numerical
approaches~\cite{Fakhari, Yarin, Bakshi, Ding3}. However, the
studies on this subject are almostly limited to the two-phase
situation~\cite{Fakhari, Bakshi, Ding3}, and the mechanism of drop
impact, especially in the case of a compound drop, has been far from
well understood. Actually, to the best of our knowledge, there is no
literature avaiable on the numerical study of a compound drop impact
on solid. To fill the gap, here we will use the phase-field-based
three-phase LB model coupled with the present wetting boundary
scheme to numerically investigate the impact of a compound drop onto
a solid circular cylinder.

\begin{figure}
\centering
\includegraphics[width=1.0in,height=1.42in]{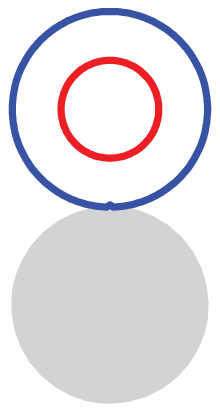}
\includegraphics[width=1.0in,height=1.42in]{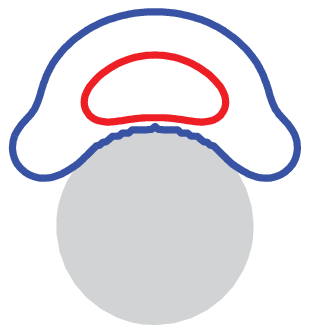}
\includegraphics[width=1.0in,height=1.42in]{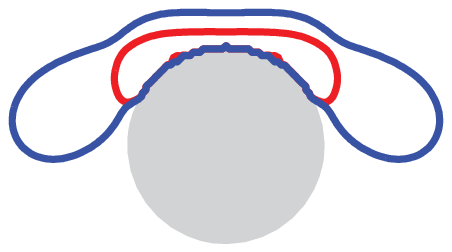}
\includegraphics[width=1.0in,height=1.42in]{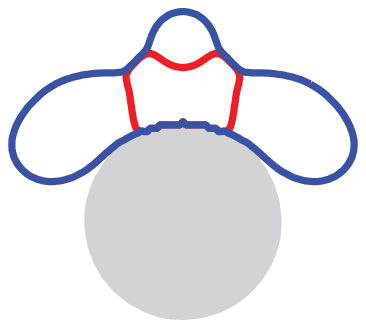}
\includegraphics[width=1.0in,height=1.42in]{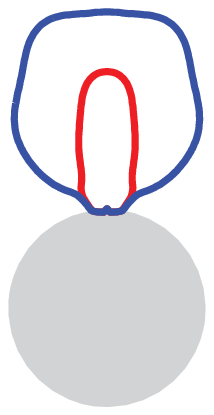}
\includegraphics[width=1.0in,height=1.42in]{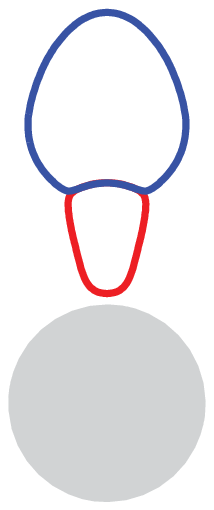}
 \tiny\caption{Snapshots of a compound drop impact on a cylindrical solid wall with $We=21.6$ and
 $\theta_{13}=150^\circ$. The dimensionless time instants from the left pattern to the right
are 0, 0.5, 2.0, 4.0, 6.0, 9.0.}
\end{figure}

\begin{figure}
\centering
\includegraphics[width=1.0in,height=1.42in]{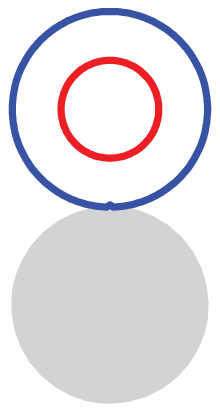}
\includegraphics[width=1.0in,height=1.42in]{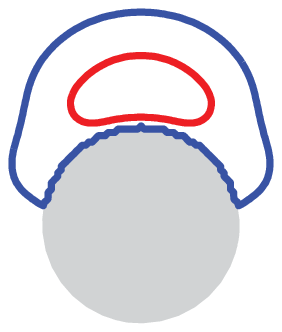}
\includegraphics[width=1.0in,height=1.42in]{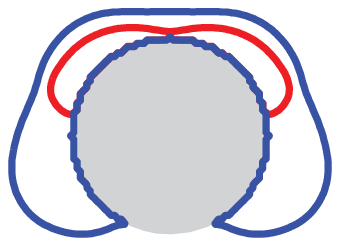}
\includegraphics[width=1.0in,height=1.42in]{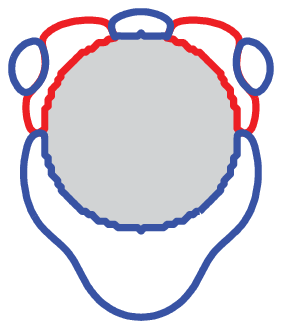}
\includegraphics[width=1.0in,height=1.42in]{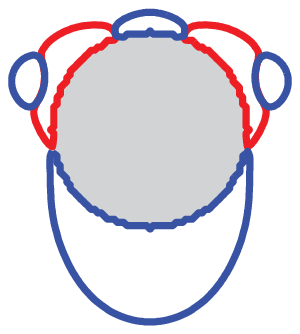}
\includegraphics[width=1.0in,height=1.42in]{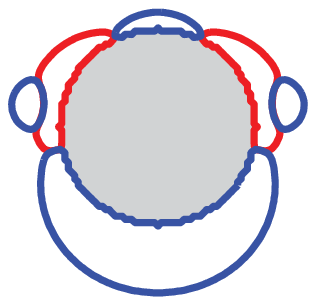}
 \tiny\caption{Snapshots of a compound drop impact on a cylindrical solid wall with $We=21.6$ and
 $\theta_{13}=30^\circ$. The dimensionless time instants from the left pattern to the right
are 0, 0.5, 2.0, 4.0, 6.0, 10.0.}
\end{figure}

The schematic of the physical problem is illustrated in Fig. 10,
where $L$ and $W$ are the length and width of the rectangular
domain, and a solid circular cylinder with the radius of $L/8$ is
centered at $(0.5L, 0.5L)$. Initially a compound drop consisting of
phases 1 and 2, in a shape of circle, is placed on the top of a
solid circular cylinder. The initial velocity in the vertical
direction with the value of $U$ is imposed for the compound drop
without the consideration of the gravity, and then it will impact
onto the solid surface. Similar to the two-phase case~\cite{Fakhari,
Ding3}, this problem can be characterized by the wall wettability in
terms of the contact angles and two group of dimensionless
parameters: the Reynolds number ($Re$) and the Weber number ($We$),
which can be defined respectively as
\begin{equation}
 Re=\frac{\xi U }{\bar{\nu}},
\end{equation}
and
\begin{equation}
 We=\frac{\bar{\rho} \xi U^2 }{\bar{\sigma}},
\end{equation}
where $\xi$ is the diameter of the compound drop given by $L/4$,
$\bar{\nu}$, $\bar{\rho}$ are the viscosity and density of the
compound drop by averaging those of two fluids, i.e.,
$\bar{\nu}=0.5(\nu_1+\nu_2)$, $\bar{\rho}=0.5(\rho_1+\rho_2)$, and
$\bar{\sigma}$ is the average surface tension computed by
$0.5(\sigma_{13}+\sigma_{23})$. Here we choose $\xi$ and $U$ as the
characteristic length and velocity, and then the characteristic time
is given by $\xi/U$.

\begin{figure}
\centering
\includegraphics[width=1.0in,height=1.42in]{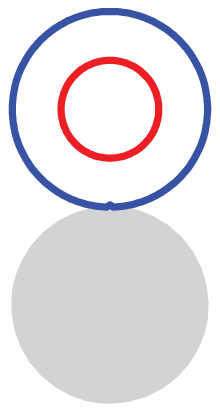}
\includegraphics[width=1.0in,height=1.42in]{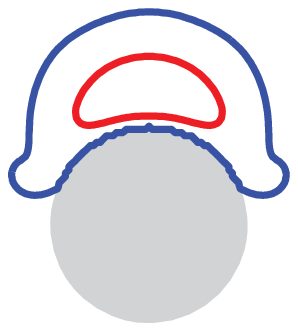}
\includegraphics[width=1.0in,height=1.42in]{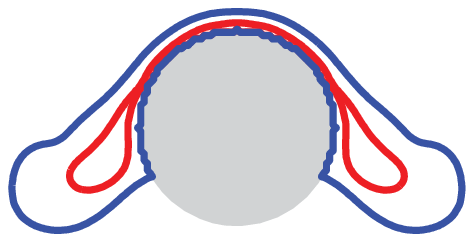}
\includegraphics[width=1.02in,height=1.42in]{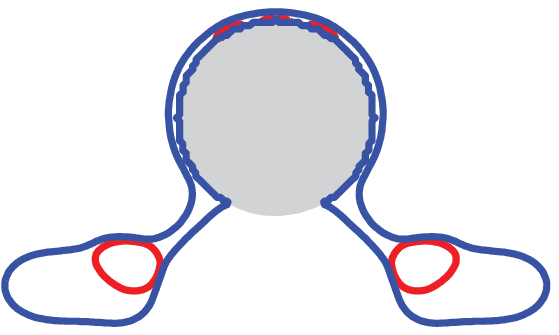}
\includegraphics[width=1.0in,height=1.42in]{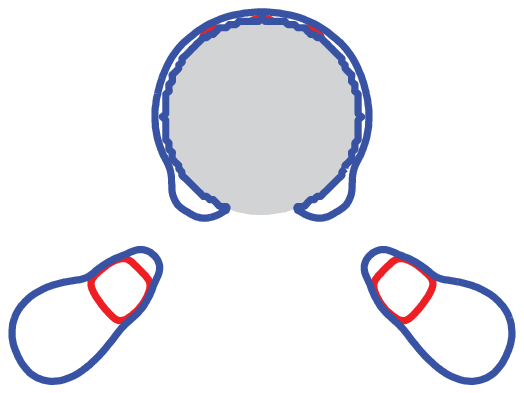}
\includegraphics[width=1.0in,height=1.42in]{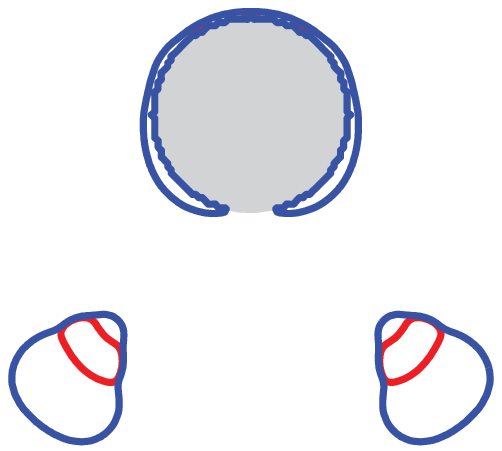}
 \tiny\caption{Snapshots of a compound drop impact on a cylindrical solid wall with $We=86.4$ and
 $\theta_{13}=30^\circ$. The dimensionless time instants from the left pattern to the right
are 0, 0.5, 2.0, 4.0, 5.0, 7.0.}
\end{figure}

\begin{figure}
\centering
\includegraphics[width=4.5in,height=3.1in]{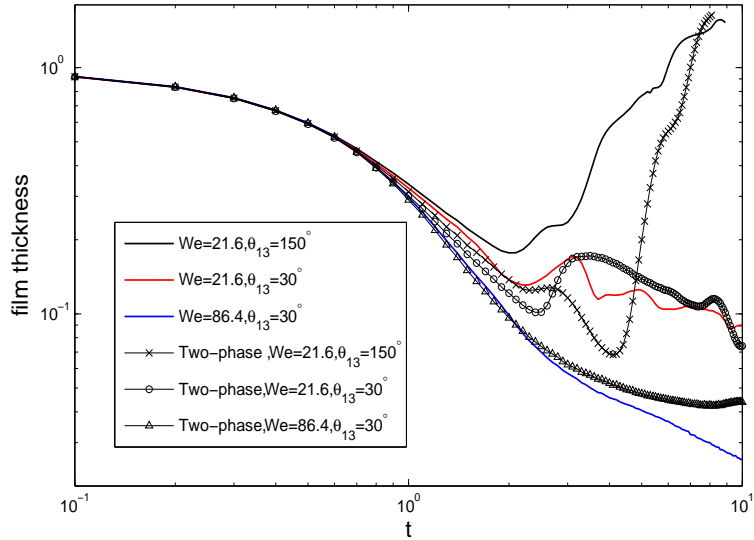}
 \tiny\caption{Temporal evolution of the normalized
film thickness ($h/\xi$) at different Weber numbers and contact
angles.}
\end{figure}

The simulation was carried out in a uniform mesh of $L\times
W=240\times280$ with periodic boundary conditions at all boundaries.
The bounce back no-slip and wetting boundary conditions are applied
on the solid surface of the circular cylinder. The profiles of the
order parameters can be initialized by
\begin{equation}
\begin{split}
 c_1(x,y)&=0.5+0.5\tanh
 \frac{2[R_1-\sqrt{(x-x_c)^2)+(y-y_c)^2}]}{D},\\
 c_2(x,y)&=0.5+0.5\tanh \frac{2[R_2-\sqrt{(x-x_c)^2)+(y-y_c)^2}]}{D}-c_1(x,y),
 \end{split}
\end{equation}
where $(x_c, y_c)=(0.5L, 0.75L)$ is the center of the compound drop,
$R_1$, $R_2$ are the radii of the compound drop and the inner drop
that are set to be $L/8$ and $L/16$. In the simulations, the
physical properties of the fluids are assumed to be
$\theta_{13}=\theta_{23}$, $\nu_1:\nu_2:\nu_3=1:1:1$, and
$\sigma_{12}:\sigma_{13}:\sigma_{23}=1:1:1$. The contact angle
$\theta_{12}$ and the Reynolds number are fixed to be $90^\circ$ and
600. Then the remaining physical parameters are given as $U=0.06$,
$\rho_1=\rho_2=10.0$, $\rho_3=1.0$, $M_1=M_2=0.1$, and
$\tau_1=\tau_2=0.8$. Here we mainly focus on the effects of the
contact angle $\theta_{13}$, and the Weber number by adjusting the
value of the surface tension. Figure 11 depicts snapshots of a
compound drop impact dynamics on a superhydrophobic cylinder with
$\theta_{13}=150^\circ$ and $We=21.6$. It can be observed that, due
to the inertia effect, the compound drop firstly spreads on the
solid cylinder, while its two-side tails are not in contact with the
surface due to the hydrophobicity property. After that, the compound
drop retracts with time and the superhydrophobicity becomes to
dominate over the inertia effect as the downward velocity is
reduced, which results in the eventual rebound of the compound drop
and completely detaches from the solid cylinder. The rebound
scenario is similar to the two-phase situation that a single drop
impacts on a flat or curve hydrophobic surface~\cite{Ding3,Yarin}.
To examine the effect of the surface wettability, we also simulated
the case with the hydrophilic surface $(\theta_{13}=30^\circ)$ and
$We=21.6$, and showed the results in Fig. 12. As can be seen from
Fig. 12 that the compound drop takes on a distinctive behaviour. It
spreads continuously along the edge of the solid cylinder, and then
covers over the whole surface after a certain time of evolution. The
compound drop goes on to move downward and portion of it hangs from
the cylinder surface. Under the action of the surface tension force,
it has mild shrinkage and finally reaches its equilibrium state.
From the perspective of each phase, we can observe that the inner
drop consisting of the 1-th phase breaks up into two symmetrical
daughter drops, while the outer drop consisting of the 2-th phase is
stretched into multiple drops, which cannot be found in the above
situation. We also consider the effect of the Weber number on the
impact dynamics. Figure 13 shows the snapshots of a compound drop
impact on a cylindrical solid wall with a large $We$ of 86.4 and
$\theta_{13}=30^\circ$. Comparatively, the compound drop also
spreads along the edge of the solid cylinder at the early time. Due
to the larger Weber number, however, the inertia pulling force
prevails over the surface tension force. Then, two thick liquid
tails are formed at the side, in addition to the liquid film around
the solid surface. Finally, the film breaks up, leading to the
release of two daughter compound drops. We should also emphasize
that the breakup phenomena of the compound drop cannot be observed
for a small Weber number.

The film thickness on the top of the solid cylinder is a concerned
physical quantity in the study of the drop impact
dynamics~\cite{Fakhari, Bakshi}. Here we also measured the film
thickness of the compound drop with different Weber numbers and
contact angles, and showed the dimensionless results in Fig. 14. Due
to the lack of the research on this subject, we cannot
quantitatively compare the present results with the available
experimental data or the numerical results. For a comparsion,
however, we presented in Fig. 14 the results of a single drop impact
on the same cylinder, which are derived from the simulations of the
present three-phase LB method by seting $c_1=0$. It can be found
from Fig. 14 that for all the cases, the film thickness of the
compound drop has a uniform curve at the initial stage, which is
also in line with that of the two-phase simulation~\cite{Fakhari}.
Then the noticeable difference between them can be observed. For the
superhydrophobic surface, the film thickness of the compound drop
continues to decrease at first and then has a rapid increase with
time, which can be attribute to that the rebound phenomenon occurs
for the compound drop. While for the hydrophilic surface, the file
thickness of the compound drop firstly has a relatively rapid
decrease and then undergoes a smooth change with time. As for the
hydrophilic surface with a larger We, the film thickness undergoes a
rapid decline with time until it reaches a residual value. In
addition, the comparison between the film curves of the compound
drop and single drop shows that they have qualitatively similar
variation tendencies under the same condition, but are
quantitatively different.

\section{Summary}\label{sec: sum}
Multiphase flows involving multiple fluid components and solid
boundary are frequently encountered in the engineering applications.
To simulate such flows, how to describe the interactions among
multiple fluids and solid surface is a crucial problem. In this
paper a suitable wetting boundary scheme that describe the
fluid-solid interaction in the framework of the lattice Boltzmann
method based on the phase-field theory is proposed. Due to the
particular choice of the wall free energy, the proposed wetting
boundary scheme can preserve the reduction consistency property with
the binary one. Coupled with this wetting boundary scheme, a lattice
Boltzmann model for three-phase flows that also satisfies
algebraical and dynamical consistency properties is used to simulate
the ternary fluid flows in contact with solid wall. Numerical
examples include the spreading of binary drops on the substrate, the
spreading of a compound drop on the substrate, and the shear of a
compound drop on the substrate. It is shown that the numerical
results of these flows agree well with the theoretical results or
some available data for a broad range of contact angles, which
provides a good validation of the present wetting boundary scheme.
As an application, we further use the lattice Boltzmann tenary fluid
model coupled with the wetting boundary condition to study a
compound drop impact on the solid cylinder, which has not been
considered before in the literature. It is found that the compound
drop dynamics can be significantly influenced by the wettability of
the cylinder surface and the Weber number, and some interesting
interfacial phenomena, including spreading, breakup, rebound, are
also observed in the simulation results. Our present discussion
focuses on the Cahn-Hilliard phase field model, and actually, the
generalized wetting boundary scheme here is appropriate for the
Allen-Cahn type phase field model. Finally, we anticipate that our
numerical method will be useful to microfluidics, material science,
and oil recovery industry.


\section*{Acknowledgments}
One of the authors (Hong Liang) would like to thank Dr. Changsheng
Huang and Dr. Chen Wu for providing insight suggestions, and this
work is also financially supported by the National Natural Science
Foundation of China(Grant Nos. 11602075, 51576079, 51406120,
11674379), and the Natural Science Foundation of Zhejiang Province
(Grant No. LY15E060007).


\end{document}